\documentclass{aa}
\usepackage{times}
\usepackage{graphics}
\usepackage{xspace}
\usepackage{epsfig}
\usepackage{jwaabib}
\usepackage{rotating}
\usepackage{dcolumn}

\begin{document}

\title{Late B-type stars and their candidate companions \\ resolved with {\em Chandra}\thanks{This work made use of observations obtained at the European Southern Observatory, La Silla, Chile (ESO program No. 67.C-0261(B)).}}
\author{B. Stelzer\inst {1} \and N. Hu\'elamo\inst {2} \and S. Hubrig\inst {2} \and H. Zinnecker\inst {3} \and G. Micela\inst {1}}

\institute{INAF - Osservatorio Astronomico di Palermo,
% INST 1
  Piazza del Parlamento 1,
  I-90134 Palermo, Italy \and
% INST 2
  European Southern Observatory, 
  Casilla 19001, Santiago 19, 
  Chile \and 
% INST 3
  Astrophysikalisches Institut Potsdam,
  An der Sternwarte 16, 
  D-14482 Potsdam,
  Germany}

\offprints{B. Stelzer}
\mail{B. Stelzer, stelzer@astropa.unipa.it}
\titlerunning{X-ray emission from B-type stars}

\date{Received $<$10-02-03$>$ / Accepted $<$13-06-03$>$}

\abstract{We present the first results from a series of {\em Chandra} 
observations carried out with the aim to examine the origin of X-ray emission 
in main-sequence late B-type stars.
X-ray detections of late-B and early A-type stars have remained a mystery 
as none of the two major theories for stellar X-ray emission applies in this spectral range:
while O- and early B-type stars drive strong winds that are subject to instabilities, 
late-type stars produce X-rays as a result of magnetic dynamo action. 
Since any dynamo works only in the presence of a convective zone, 
early-type stars are not 
magnetically active. We use high spatial resolution X-ray observations
to enlighten the prevalent 
speculation that previously unknown late-type or low-mass companion stars are 
the sites of the X-ray
emission, instead of the B-type primaries. Here we present the results 
for HD\,1685, HD\,113703, HD\,123445, HD\,133880, and HD\,169978. 
Adaptive optics observations have recently revealed at least one 
faint object near each of these B-type stars (at separation of $1-6^{\prime\prime}$). 
Four of the new infrared objects show infrared colors and magnitudes typical for low-mass 
pre-main sequence stars, and are likely true companions to the $\sim 10-50$\,Myr old B-type 
stars. 
These multiple systems are now resolved for the first time in X-ray light. 
We uncover that four of the new companions are X-ray
emitters, and the fifth one is likely to be a weak X-ray source below the detection limit. 
Three of the B-type primaries are X-ray dark down to the detection limit of 
$L_{\rm x} \sim 10^{28}$\,erg/s. But we {\em do} detect X-ray emission from the position 
of HD\,1685\,A and HD\,169978\,A. 
The latter one indeed is a spectroscopic binary. 
The characteristics of all X-ray sources are compatible with
those of typical young late-type stars: 
hard X-ray spectrum ($kT > 0.5$\,keV) and high X-ray luminosity ($\lg{L_{\rm x}} \sim 29...30$\,erg/s). Spectroscopic observations in the infrared 
will solve the question whether the one remaining
X-ray detected B-star in our sample, HD\,1685\,A, 
also has an even closer companion or whether this is an intrinsic X-ray emitter. 
\keywords{X-rays: stars -- stars: early-type, late-type, coronae, activity}
}

\maketitle

\section{Introduction}\label{sect:intro}

X-ray observations performed by the {\em Einstein} and {\em ROSAT}
missions have revealed that X-rays are emitted by stars throughout the
Hertzsprung-Russell diagram (e.g. \cite{Vaiana81.1}, 
\cite{Schmitt95.1}, \cite{Neuhaeuser95.1}, \cite{Huensch99.1}). 
For stars on the main-sequence (MS) two mechanisms are known to be responsible for the
observed emission. 

In hot stars the X-rays are produced
by instabilities arising in the strong radiatively driven stellar
winds (\cite{Lucy80.1}). 
O-type stars are characterized by a scaling between X-ray and 
bolometric luminosity of $L_{\rm X}/L_{\rm bol} \approx 10^{-7}$ 
(e.g. \cite{Berghoefer97.1}). 
This empirical relation can be reproduced by models for optically thick
winds that take account of X-ray attenuation (\cite{Owocki99.1}).
However, the observed correlation breaks down near spectral types B2, where the 
$L_{\rm X}/L_{\rm bol}$ ratio 
falls by more than one order of magnitude, and stars cooler than B4 require
wind filling factors larger than unity, i.e. their X-ray emission 
can not be reconciled with any wind model (\cite{Cohen97.1}). 

In late-type stars a
solar-like magnetic dynamo driven by rotation and convection 
is thought to produce the observed (X-ray)
activity (\cite{Parker55.1}, \cite{Parker93.1}, \cite{Ruediger95.1}). 
Interior models based on mixing-length theory place the 
transition from radiative to convective envelope near $T_{\rm eff} \sim 8300$\,K 
(e.g. \cite{ChristensenDalsgaard00.1}). 
However, the minimum depth of a convective
envelope able to support magnetic activity is not well
established. Based on X-ray observations the onset of significant 
coronal emission is placed somewhere between spectral type A7 and F4
(\cite{Schmitt85.1}, \cite{Schmitt97.1}). Spectroscopic observations
in the ultraviolet (UV) 
seem to indicate that chromospheres exist in most early-F type stars
(\cite{Simon91.1}, \cite{Simon94.1}). 
Furthermore, a survey in the far-UV among A-type stars with {\em FUSE} has 
revealed that chromospheric line fluxes are similar in strength to the Sun for stars
with effective temperature $\leq 8200$\,K (corresponding to spectral
type $\sim$\,A4 on the MS), 
while above this temperature activity drops abruptly (\cite{Simon02.1}). 

Stars with spectral types late B and early A 
do not drive strong enough winds nor do they possess
convective zones necessary to sustain a magnetic dynamo.
Consequently no X-ray emission is expected. Nevertheless,
the X-ray detection of these stars has been reported in several works
(e.g. \cite{Caillault89.1}, \cite{Micela90.1}, \cite{Grillo92.1},
\cite{Schmitt93.2}, \cite{Berghoefer94.1}, \cite{Caillault94.1},
\cite{Stauffer94.1}, \cite{Simon95.1}, \cite{Berghoefer96.1},
\cite{Panzera99.1}, \cite{Huelamo00.1}, \cite{Stelzer01.1},
\cite{Daniel02.1}, \cite{Briggs03.1}).
Lacking another explanation, the X-ray emission of these 
stars is commonly attributed to unresolved late-type companions.
Because a large fraction of the X-ray detected late B-type stars
belong to rather young ($\sim 10^{7...8}$\,yrs) 
stellar groups (e.g., Sco\,OB2, Carina-Vela, Tucanae),
most of the predicted unresolved late-type stars may be young stars still
contracting to the MS or just arrived on the zero-age MS, 
if bound to the primaries.  This idea has been
supported by (i) the high X-ray luminosities of the late B-type stars,
comparable to those of pre-MS stars (Bergh\"ofer et al. 1997) and
(ii) the spectral distribution of their X-ray emission which is
similar to those of young late-type stars, i.e., they are hard X-ray
emitters (Hu\'elamo et al. 2000).

In order to check the hypothesis of unresolved companions to late
B-type stars \citey{Berghoefer94.1} carried out 
{\em ROSAT} High Resolution Imager (HRI) X-ray observations   
of visual binaries with separations $>\,10^{\prime\prime}$, i.e. those
clearly resolvable by the HRI. In these observations only in $1$ out of $8$ 
cases the X-ray emission could be ascribed to the known visual
late-type companion. On the other
hand, {\em ROSAT} HRI studies of visual binary systems comprised
of early-type stars and post-T Tauri stars (also known as Lindroos
systems), has shown that both the late B-type primaries and their
late-type companions emit X-rays at similar levels 
(\cite{Schmitt93.2}, \cite{Huelamo00.1}).  The similarity of the X-ray
properties of the Lindroos primaries and secondaries supports the
hypothesis that the X-ray emission from the late B-type stars in fact 
originates from closer pre-MS late-type companions unresolvable by the
{\em ROSAT} HRI.

In view of the large number of X-ray
detected late B- and early A-type stars and the inconclusive results of
the existent observations, the problem clearly needs further
attention. Clarification of this issue can be obtained by carrying
out high spatial resolution X-ray observations to precisely 
locate the X-ray source. The exceptional spatial resolution of {\em Chandra} 
allows to push closer and closer in: systems as close as $\sim 1^{\prime\prime}$
can now be studied. At the distance of $100-200$\,pc where many 
of the X-ray emitting B- and A-type stars are located, this corresponds to 
$< 200$\,AU separation. This is much smaller than the maximum separation of
visual binaries in the solar neighborhood 
(see \cite{Close90.1}, \cite{Duquennoy91.1}), and thus the
systems are not unlikely to be bound. 

We have started a series of {\em Chandra} observations pointing at selected
multiple stars with a B-type primary. Here we report on the first results.  
Five X-ray emitting late B-type stars with recently identified faint objects 
at separations between $\sim 1-6^{\prime\prime}$ were targeted with
the Advanced CCD Imaging Spectrometer. This instrument, in addition to
unsurpassed spatial resolution, provides spectral
capabilities useful to constrain the nature of the X-ray emitter. 

In Sect.~\ref{sect:sample} we explain the selection of the sample
observed with {\em Chandra}.  In Sect.~\ref{sect:observations} we
describe the observations and the data analysis. All detected X-ray sources
related with components from the B-star sample are presented in
Sect.~\ref{sect:detections}.  The X-ray spectra and luminosities are
discussed in Sect.~\ref{sect:spectra} and Sect.~\ref{sect:lx}.  We
examine the nature of the companions (Sect.~\ref{sect:cmd}), and
discuss the X-ray characteristics of wide ($>10^{\prime\prime}$)
companions in the Lindroos systems among our targets
(Sect.~\ref{sect:lindroos}). 
The results are discussed in Sect.~\ref{sect:discussion}.

\section{Sample selection}\label{sect:sample}

The sample of {\em Chandra} targets is based on near-infrared (IR) adaptive optics
(AO) observations with ADONIS performed 
by \citey{Hubrig01.1} and \citey{Huelamo01.1}. Both of these
studies aimed at detecting new late-type companions to X-ray detected 
late B-type stars, in search for the origin of their X-ray emission.

\citey{Hubrig01.1} carried out diffraction limited near-IR observations 
of $49$ X-ray emitting late B-type stars
extracted from the {\em ROSAT} study of \citey{Berghoefer96.1}. 
As a result they reported the discovery of new
companions to $19$ of their X-ray selected stars, with
separations from the B-type primary between
$0.2-14^{\prime\prime}$. 

In an analysis of X-ray emission from Lindroos binaries \citey{Huelamo00.1}
found that three of the X-ray detected B-type primaries show X-ray properties
very different from those of the other early-type components in the Lindroos
sample, suggestive of further unknown late-type companions. 
Subsequently \citey{Huelamo01.1} 
performed AO observations on the B-type primaries in Lindroos systems 
and identified faint objects near one of them. 

We selected a homogeneous subgroup 
of the \citey{Hubrig01.1} and \citey{Huelamo01.1}
samples for observations with {\em Chandra}. The first group of objects 
consists of four systems: HD\,1685, HD\,123445, HD\,133880, and HD\,169978. 
We add an archived {\em Chandra} observation of the Lindroos system HD\,113703 
for which we also found a likely late-type companion with ADONIS 
(Hu\'elamo et al., in prep.). The same faint IR object has been detected by
\citey{Shatsky02.1}. 

%
% Position + distance of primary from Hipparcos data base
% Sptype from Berghoefer et al (1996)
% Separation and Position angle from Huelamo et al (2001), Hubrig et al. (2001),
%   and Shatsky & Tokovinin (2002)
% 
\begin{table*}
\begin{center}
\caption{Target list of {\em Chandra} observed B-star systems: optical parameters of primaries, separation and position angle of newly discovered component in the system, and information about the {\em Chandra} observations. Only the companions newly identified with AO observations are listed. Two of the stars are Lindroos systems, i.e. in addition they have previously known wider companions which are discussed in Sect.~\ref{sect:lindroos}. HD\,123445 has two faint companions discovered with ADONIS.} 
\label{tab:obslog}
\begin{tabular}{lrrrlrrrrr}
\noalign{\smallskip} \hline \noalign{\smallskip} 
Designation & \multicolumn{2}{c}{Position$^{(1)}$}                                       & Dist$^{(1)}$ & SpT$^{(2)}$ & \multicolumn{1}{c}{$\lg{L_{\rm bol,A}^{(3)}}$} & \multicolumn{1}{c}{Sep$^{(4)}$}                 & \multicolumn{1}{c}{PA$^{(4)}$}   & \multicolumn{2}{c}{ACIS observations$^{(5)}$} \\
            & \multicolumn{1}{c}{$\alpha_{2000}$} & \multicolumn{1}{c}{$\delta_{2000}$}  & [pc]         &             & [erg/s]                                        & \multicolumn{1}{c}{[$^{\prime\prime}$]} & \multicolumn{1}{c}{[$^\circ$]} & ObsID  & Obs.Date  \\
\noalign{\smallskip} \hline \noalign{\smallskip} 
HD\,1685        & 00 20 39.0 & $-$69 37 29.7 &  94 & B9   & $35.42$ & 2.28  & 211.4 & $2541$ & Sep 23, 2002 \\
\noalign{\smallskip} \hline \noalign{\smallskip} 		     
HD\,113703      & 13 06 16.7 & $-$48 27 47.8 & 127 & B5   & $36.26$ & 1.551 & 268.2 & $0626$ & Jun 10, 2000 \\
\noalign{\smallskip} \hline \noalign{\smallskip} 		     
HD\,123445      & 14 08 51.9 & $-$43 28 14.8 & 218 & B9   & $35.84$ & 5.56/5.38 & 65.0/64.0 & $2542$ & Jan 06, 2002  \\
\noalign{\smallskip} \hline \noalign{\smallskip} 		     
HD\,133880      & 15 08 12.1 & $-$40 35 02.1 & 126 & B8   & $35.76$ & 1.222 & 109.2 & $2543$ & Apr 04, 2002 \\
\noalign{\smallskip} \hline \noalign{\smallskip} 		     
HD\,169978      & 18 31 22.4 & $-$62 16 41.9 & 147 & B7   & $36.29$ & 3.085 & 168.7 & $2544$ & Jun 21, 2002 \\
\noalign{\smallskip} \hline \noalign{\smallskip} 
\multicolumn{10}{l}{$^{(1)}$ {\em Hipparcos} position and distance for the B-type star,} \\
\multicolumn{10}{l}{$^{(2)}$ spectral types adopted from \protect\citey{Berghoefer96.1} who originally extracted them from the Yale Bright Star Catalogue,} \\
\multicolumn{10}{l}{\hspace*{0.35cm} (see \protect\cite{Hoffleit91.1});} \\
\multicolumn{10}{l}{$^{(3)}$ bolometric luminosities of the B-type star were derived from the $V$ magnitude using the bolometric corrections by \citey{Schmidt-Kaler82.1};} \\
\multicolumn{10}{l}{$^{(4)}$ separation and position angle from \protect\citey{Hubrig01.1}, \protect\citey{Shatsky02.1}, and \protect\citey{Huelamo01.1};} \\
\multicolumn{10}{l}{\hspace*{0.35cm} separations have been measured in the ADONIS detector space;} \\
\multicolumn{10}{l}{$^{(5)}$ Obs-ID\,$0626$ was performed with ACIS-S, all other observations with ACIS-I.} \\
%\multicolumn{10}{l}{$^{(6)}$ this star has two IR-companions termed 'C' and 'D' (see \protect\cite{Huelamo01.1})} \\ 
\end{tabular}
\end{center}
\end{table*}

The selected group of stars fullfills the following selection criteria:

1. The stars have previously unknown visual companions revealed by 
AO observations. 

2. The separations between the new companions and the
late B-type star are larger than $1^{\prime\prime}$, that is 
clearly resolvable by {\em Chandra}, but smaller than $6^{\prime\prime}$. 
This means we studied the X-ray emission from sources that could not be 
resolved by previous instruments.
Note that systems with
separation $>\,5^{\prime\prime}$ should in principle be resolvable by
the {\em ROSAT} HRI. However, the {\em ROSAT} HRI failed in resolving 
some of the Lindroos systems with separations between
$5-10^{\prime\prime}$ (\cite{Huelamo00.1}).
 
3. The primaries do not show signs of intrinsic binarity
according to the {\em Hipparcos} data base and the 
$\Delta\mu$ database (\cite{Wielen00.1}).  
%and their published spectroscopic data. 
This way we minimize the chance that any
X-ray emission discovered by {\em Chandra} at the position of the
B-type star is due to a very close late-type companion not
resolvable with both the AO IR images and {\em Chandra}.
One of our targets is however a spectroscopic binary
(\cite{Aerts99.1}).

To summarize, the sample we present in this paper is composed of five
B-type stars with recently identified faint IR objects close-by.  The
{\em Hipparcos} position of our targets, their distance, spectral type,
and bolometric luminosity are listed in Table~\ref{tab:obslog}.
We give also the separation and position
angle of the ADONIS companions. Two of the {\em Chandra} targets
(HD\,113703 and HD\,123445) are Lindroos systems with an additional
previously known late-type companion.  The separations for
these secondaries are large ($> 11^{\prime\prime}$), and not of
interest for the main aim of our {\em Chandra} study, isolating X-rays
from the primary and the newly discovered ADONIS companions. 
But one of these Lindroos secondaries was not resolved in X-rays
before, and therefore we discuss the X-ray properties of the
late-type Lindroos stars in Sect.~\ref{sect:lindroos}.

We point out that for the moment it remains unclear whether
the newly discovered IR objects are physically bound to the B-type stars. 
Confirmation that 
they are true companions requires observations of their proper motion and/or spectra. 
The B-type stars in our sample are on the MS. A small doubt remains only for
HD\,169978, which has luminosity class III according to 
SIMBAD. 
%\footnote{The SIMBAD data base is accessible through the URL at http://simbad.u-strasbg.fr/Simbad.}. 
But in Sect.~\ref{sect:cmd} 
we show that this star is more likely to be on the MS. The contraction timescale
of pre-MS stars to the MS is comparable to the mean lifetime of late B-type stars
on the MS. Hence, if the IR sources are bound to the B-type star they must be
young late-type stars in approach to the MS. In that case 
optical spectroscopy should reveal a Li\,I absorption feature at
$6708$\,\AA~ 
indicating their pre-MS nature. Lacking definite information about their status
we will for simplicity continue calling the IR objects 'companions', and the B-type stars
'primaries'. The issue is discussed in more detail in 
Sects.~\ref{sect:cmd} and~\ref{sect:lindroos}.

\section{Observations and Data Analysis}\label{sect:observations}

All stars introduced in Sect.~\ref{sect:sample} were observed
with {\em Chandra} using the Advanced CCD Imaging Spectrometer (ACIS)  
in imaging mode. 
The Obs-ID and date of all {\em Chandra} observations can be found 
in Table~\ref{tab:obslog}. 

Except for HD\,113703 ACIS-I was used as the prime instrument because 
of the optical brightness 
of the B-type stars ($V \sim 5...6\,$mag), which is slightly below the limiting 
magnitude of ACIS-S. 
Two of the CCDs of the ACIS-S array were also turned on, but their data will not 
be discussed here.
The somewhat lower sensitivity of ACIS-I with respect to ACIS-S did not restrict 
our observations, because the objects are bright X-ray sources.
The net exposure time per target was between $2300$\,s and $2400$\,s.

The observation of HD\,113703 was performed with the spectroscopic array of ACIS. 
Only the two central chips (S2 and S3) were turned on. The frame time had
been reduced to $0.9$\,s. Generally this is useful in the case of X-ray bright
targets to avoid pile-up. To enhance the observing
efficiency despite the small frame time the $1/4$ subarray of the chips were
used. This observation was much longer than the other ones, $\sim12$\,ksec. 
The different instrument setup
and exposure time for HD\,113703 stem from the fact that we added this
observation from the archive to the projected sample.

The data analysis was carried out using the CIAO software 
package\footnote{CIAO is made available by the CXC and can be downloaded 
from \\ http://cxc.harvard.edu/ciao/download-ciao-reg.html} version 2.3
in combination with the calibration database (CALDB) version 2.18.
We started our analysis with the level\,1 events file provided by the
pipeline processing at the {\em Chandra} X-ray Center (CXC).
Observation $0626$ was processed at the CXC with CALDB version 2.3, and since
it involves the S3 chip of ACIS we had to apply a new gain map and updates 
on the geometry (focal length, ACIS pixel size and chip positions). 
All other observations discussed here have been processed with CALDB version 2.9
or later where these modifications were performed automatically.  
In the process of converting the level\,1 events file to a level\,2 events file
for each of the observations we performed the following steps: 
We filtered the events file for event grades
(retaining the standard {\em ASCA} grades $0$, $2$, $3$, $4$, and $6$), 
and applied the standard good time interval (GTI) file. 
Events flagged as cosmic ray afterglow 
were retained after inspection of the images revealed that a substantial
number of source photons erroneously carry this flag. 
We removed the pixel randomization which is automatically applied by the CXC pipeline. 
Pixel randomization deteriorates the spatial resolution. 
Since the positional accuracy is particularly
important to our observations we also checked the astrometry for any known 
systematic aspect offsets using CIAO software. In three cases 
(Obs-ID 0626, Obs-ID 2542, and Obs-ID 2543) 
we found that a small aspect correction is needed.
We took care of this by modifying the respective header keywords in the events
level\,2 file.

\subsection{Source Detection and Identification}\label{subsect:srcdet_and_iden}

Source detection was carried out with the {\it wavdetect} algorithm (\cite{Freeman02.1}).
This algorithm correlates the data with a mexican hat function
to search for deviations from the background. The {\it wavdetect} 
mechanism is well suited for separating closely spaced point sources.  
We used wavelet scales between $1$ and $8$ in steps of $\sqrt{2}$. 
The source detection was performed on an unbinned image, to achieve the 
best-possible spatial resolution. For the ACIS-I observations  
we used images with size of 
$2048 \times 2048$ pixels centered on the {\em Hipparcos} position of the primary. 
For the ACIS-S observation we examined only the S3 chip, i.e. the image size
was $1024 \times 1024$ pixels. 
The threshold for the significance of the detection was set to $2\,10^{-7}$.
For this value the detection of one spurious source is expected in a 
$2048 \times 2048$ pixel wide image. 
Although our targets are always located in the center of the ACIS field
we analysed the full image. This way other X-ray sources can be used to 
cross-check the positional accuracy.

With the threshold given above we detected between $10$ and $16$
sources per field.  
In all cases at least one X-ray source is detected
near our target, and these sources are always among the brightest in
the respective {\em Chandra} field.  For each of the exposures we
measured the offset between the X-ray detections and the B-type star
and the offset between the X-ray detections and the newly discovered AO 
companions listed in Table~\ref{tab:obslog}. Then for each of the X-ray
sources corresponding to any one of the components of our targets
photons were extracted from the $3\,\sigma$ source ellipse provided by
{\it wavdetect}.
For the X-ray undetected components in our sample 
we computed upper limits following the
prescription for Poisson-distributed counting data given by \citey{Kraft91.1}. 

To check the accuracy of the satellite's aspect solution 
we cross-correlated all detected X-ray sources with optical catalogues 
({\em Guide Star Catalogue} [GSC] and {\em US Naval Observatory Catalogue} [USNO\,A2.0]) 
using an error radius of $2^{\prime\prime}$. The pointing
accuracy is expected to be better than this value. 
Most of the X-ray sources do not have a known optical counterpart within this 
search radius. Since many of them are very faint ($<10$ counts) 
this may indicate that they are spurious detections or could be 
as yet unknown extragalactic objects.  
Although we chose the detection significance threshold such that only one 
spurious source is expected per image previous
observations have shown that the actual number of them is larger; see e.g.
\citey{Daniel02.1}. 
Despite this fact each of the fields contains at least two
X-ray sources with optical counterpart. Inspection of the offsets between
optical and X-ray position 
for these sources shows no systematic effect, and we conclude that no aspect
correction is required (besides the correction described in the previous
section).

\section{X-ray Detections}\label{sect:detections}

\begin{figure*}
\begin{center}
\parbox{18cm}{
\parbox{6cm}{
%\resizebox{6cm}{!}{\includegraphics{/um1/stelzer/chandra/bstars/200149/reg/ds9_hd1685_xray_tiny.ps}}
\resizebox{6cm}{!}{\includegraphics{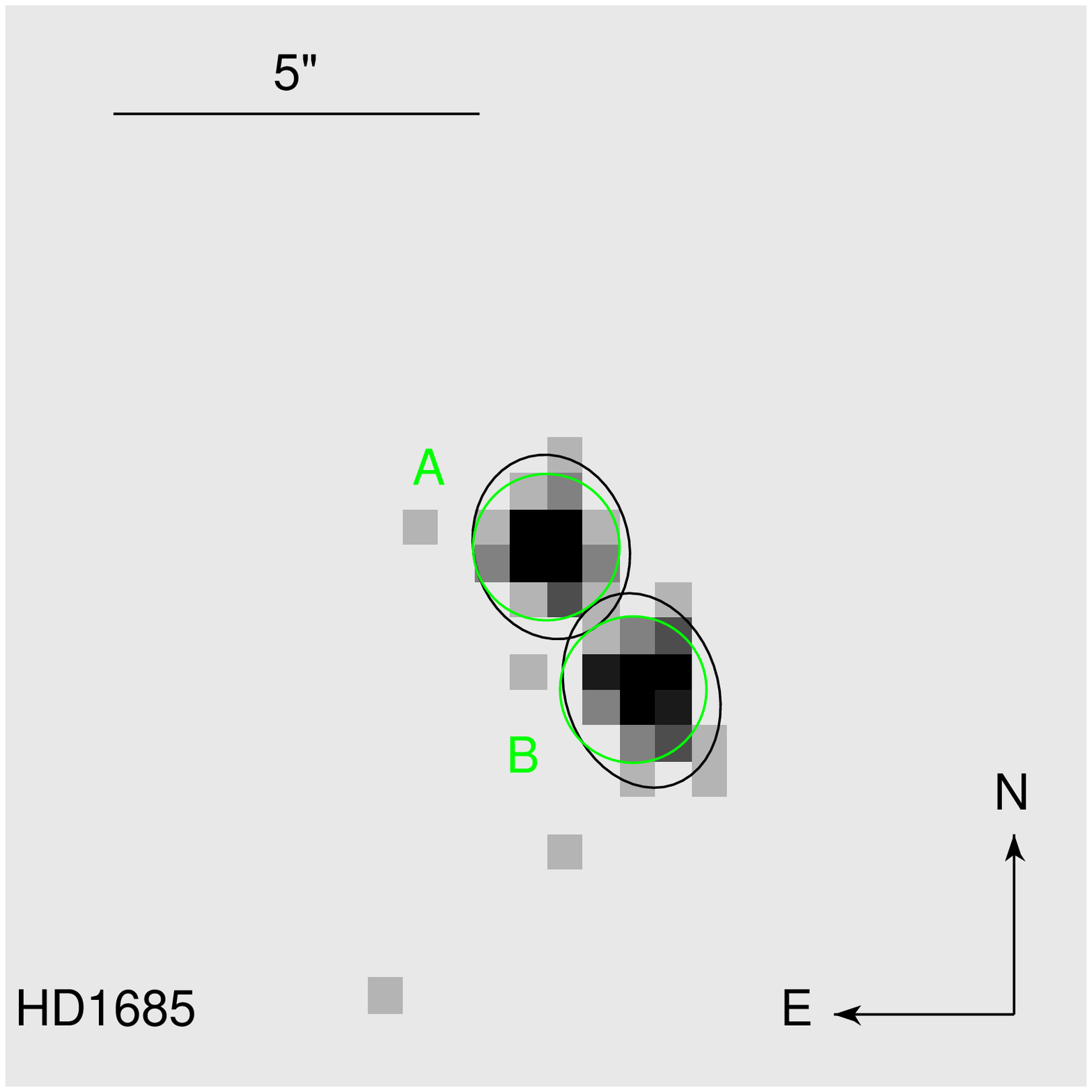}}
}
\parbox{6cm}{
%\resizebox{6cm}{!}{\includegraphics{/um1/stelzer/chandra/bstars/200051/reg/ds9_hd113703_xray_small.ps}}
\resizebox{6cm}{!}{\includegraphics{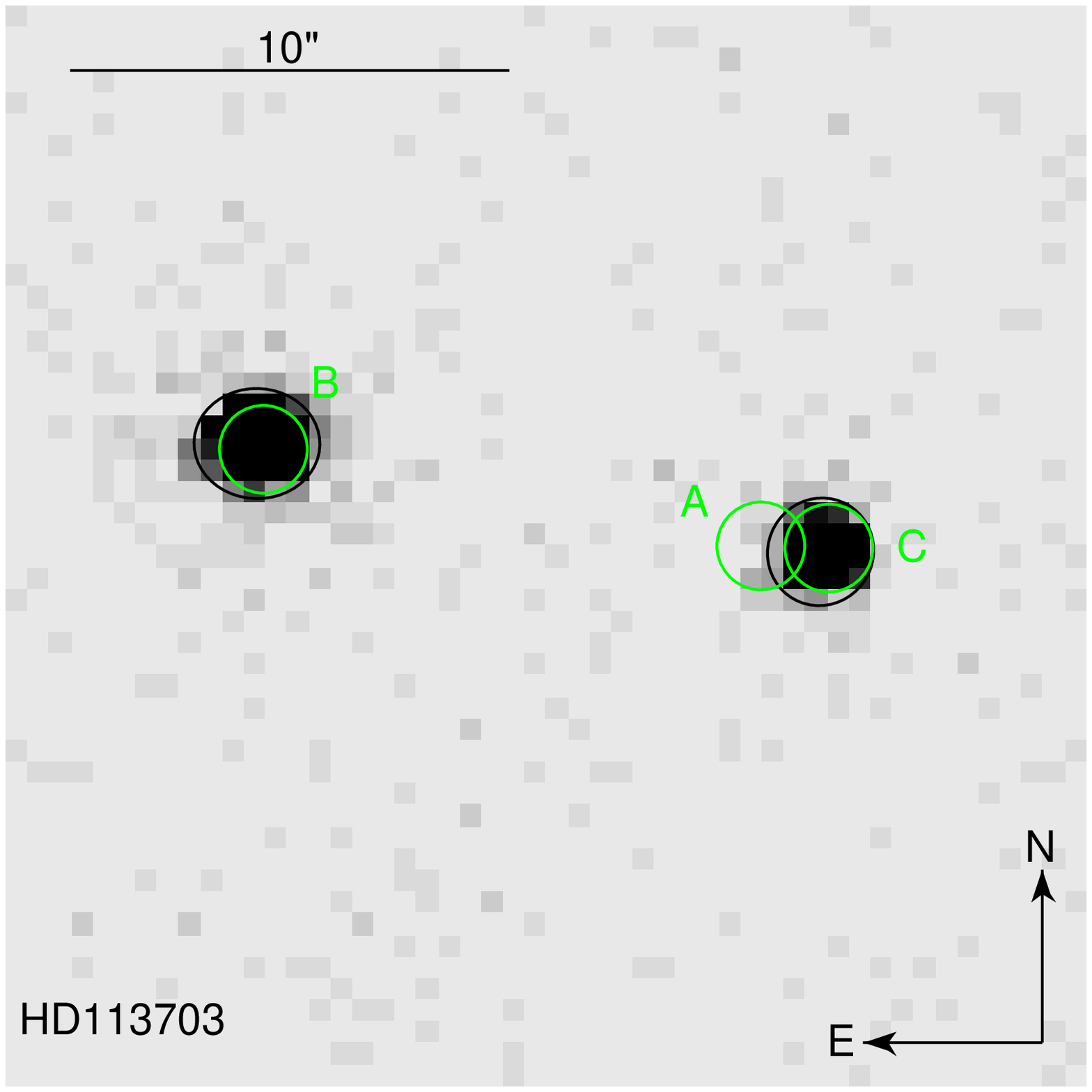}}
}
\parbox{6cm}{
%\resizebox{6cm}{!}{\includegraphics{/um1/stelzer/chandra/bstars/200150/reg/ds9_hd123445_xray_tiny.ps}}
\resizebox{6cm}{!}{\includegraphics{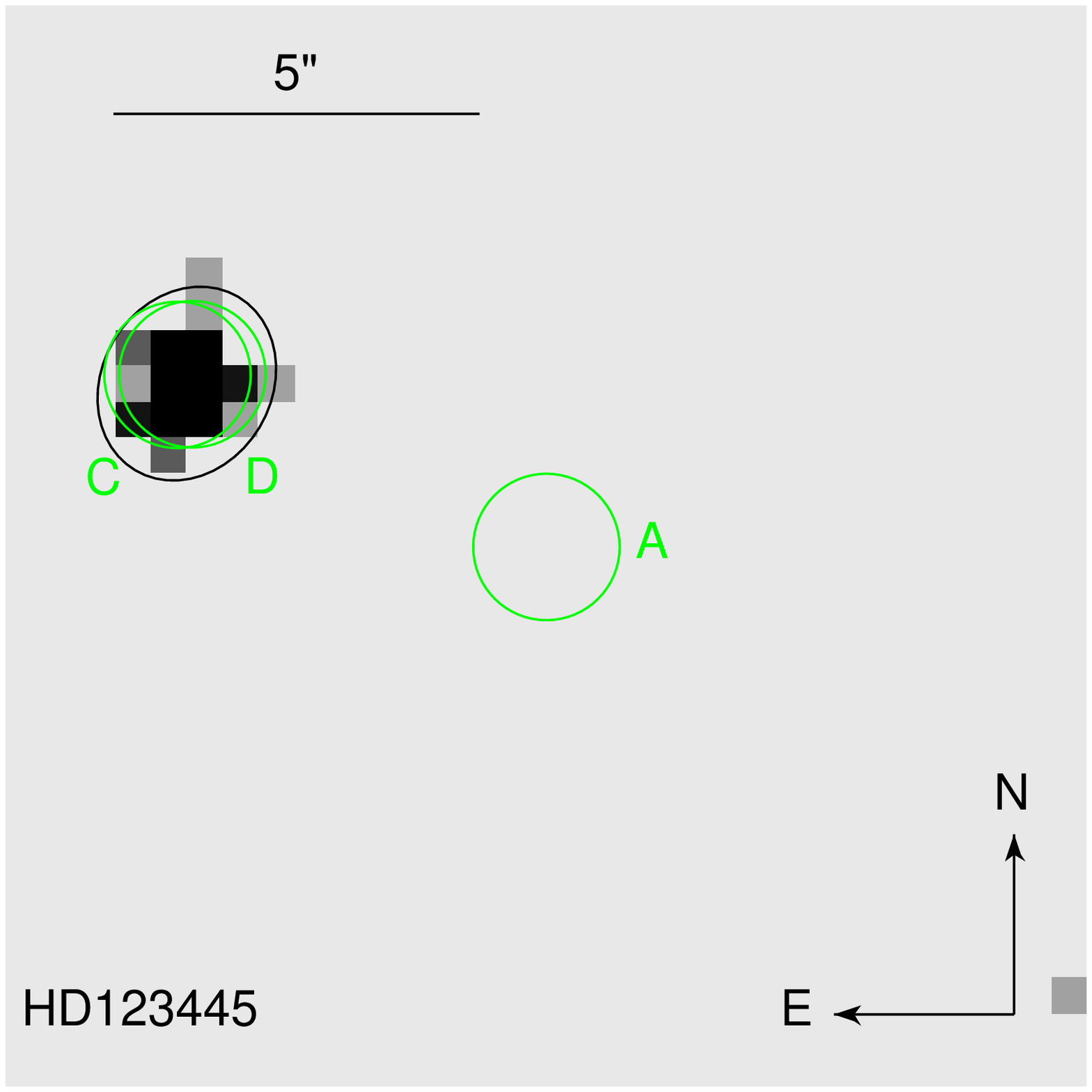}}
}
}
\parbox{18cm}{
\parbox{6cm}{
%\resizebox{6cm}{!}{\includegraphics{/um1/stelzer/chandra/bstars/200151/reg/ds9_hd133880_xray_tiny.ps}}
\resizebox{6cm}{!}{\includegraphics{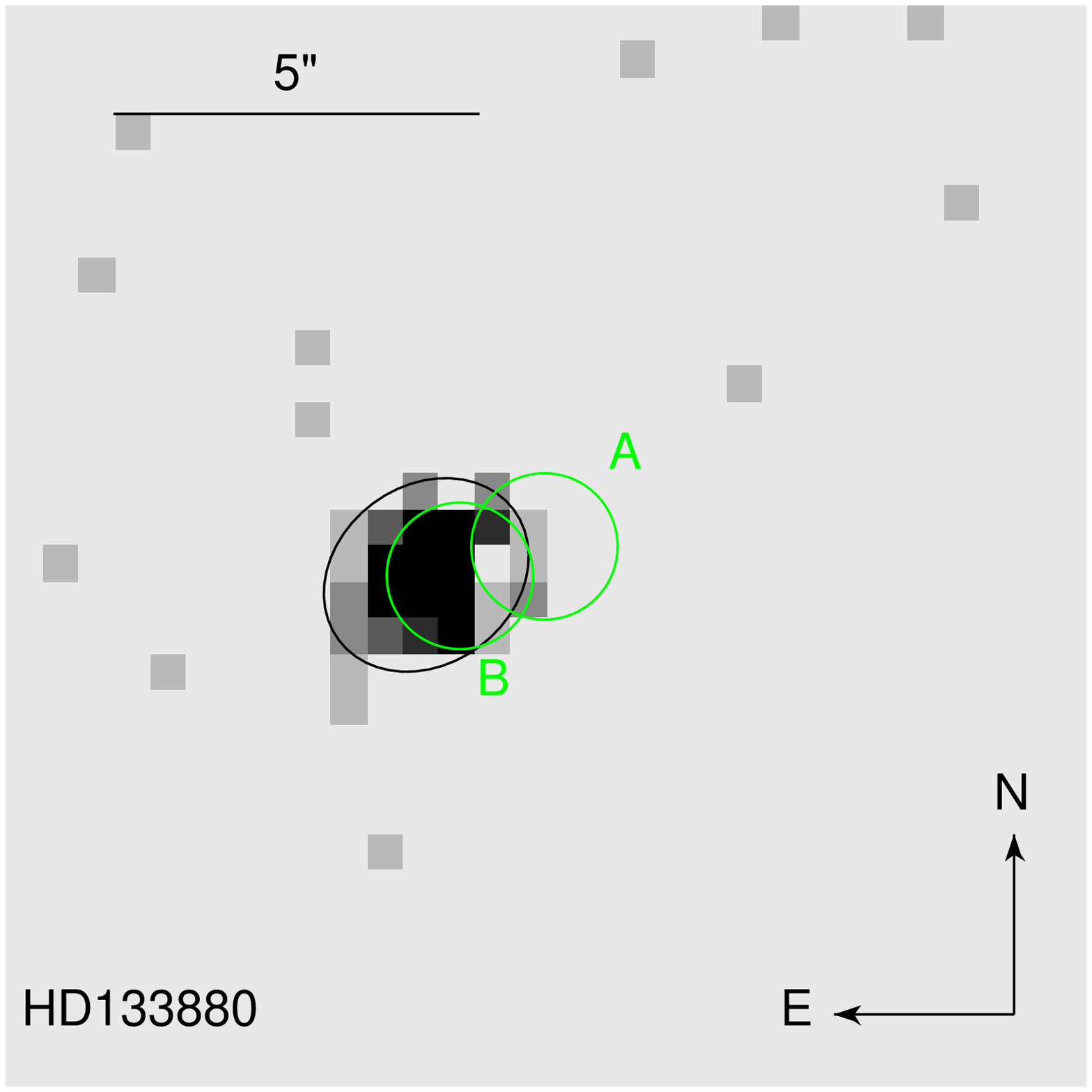}}
}
\parbox{6cm}{
%\resizebox{6cm}{!}{\includegraphics{/um1/stelzer/chandra/bstars/200152/reg/ds9_hd169978_xray_tiny.ps}}
\resizebox{6cm}{!}{\includegraphics{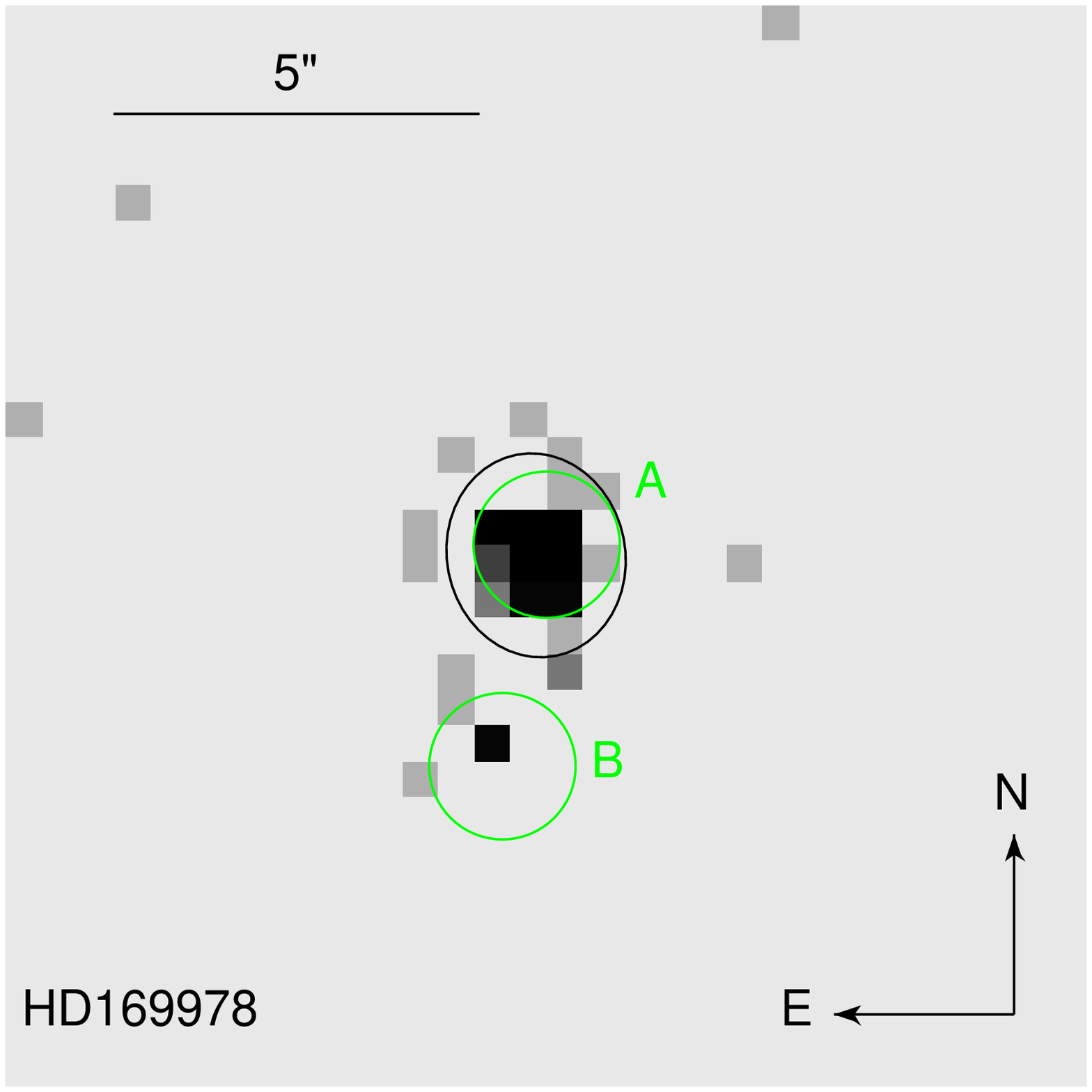}}
}
\parbox{6cm}{\caption{
{\em Chandra} ACIS images of HD\,1685,  HD\,113703, HD\,123445, HD\,133880, and HD\,169978. The size of the images is $30 \times 30$~pixel, except for HD\,113703 where it is $50 \times 50$~pixel to include the Lindroos companion HD\,113703\,B at a separation of $\sim 11^{\prime\prime}$. Grey circles denote the optical/IR position of the individual components in the multiple system ($1^{\prime\prime}$ radius). Black ellipses mark the X-ray sources detected with {\em wavdetect}.}\label{fig:acis_images}}
}
\end{center}
\end{figure*}

Fig.~\ref{fig:acis_images} shows the central portion of the ACIS images around our targets.
We overplot the source extraction area computed with {\it wavdetect} (dark ellipses),
as well as the {\em Hipparcos} position of the primary and the IR position of the 
companions (grey circles). Only in one case (HD\,1685) two
X-ray sources are detected, i.e. both B-type star and IR object are bright 
X-ray emitters. In four of the five 
targets an X-ray source can be associated with the new ADONIS object.  

HD\,169978 is the only case in the sample studied here where there is clearly
no X-ray source at the position of the nearby companion. However, 
we find a total of $6$ counts at the position of this object 
versus $<1$~count on average in a source-free region of the same area. 
This suggests that HD\,169978\,B is an X-ray source below the detection limit. 
In Sect.~\ref{sect:lx} 
we estimate an upper limit to the X-ray luminosity of this object. 

The field of HD\,169978  
contains one X-ray source which has similar brightness to HD\,169978\,A. 
This X-ray source is identified with HD\,170046, an A8/9 IV
type star according to SIMBAD, already detected with the {\em ROSAT} HRI. 
%\footnote{The SIMBAD data base is accessible through the
%URL at http://simbad.u-strasbg.fr/Simbad.} 
The offset between the X-ray and the optical position in the Chandra image 
is $1.4^{\prime\prime}$. A translational position error of this order 
is not ruled out. However, in view of the high precision of the X-ray coordinates
in the other {\em Chandra} fields examined here we consider this an unlikely
possibility. The optical position is NW of the X-ray
source, i.e. it points opposite to the position angle of the AO companion of
HD\,169978\,A. 
Therefore,  if the offset between HD\,170046 and the adjacent X-ray source 
was an error in the aspect solution of 
{\em Chandra} the X-ray source near HD\,169978 would remain unexplained, as it would
then coincide neither with the primary nor the secondary of this system. 
We speculate that HD\,170046 may be another 
candidate for an X-ray emitting A-type star with a possible    
late-type companion separated by $\sim 1.5^{\prime\prime}$. 
Recall, however, that late A-type stars may possess shallow convection zones, and
thus the X-ray emission could also be intrinsic to HD\,170046. 
In terms of hardness ratio HD\,170046 resembles the other 
sources discussed in this paper, $HR=0.21 \pm 0.15$ for $S=0.5...1.0$\,keV and
$H=1.0-8.0$\,keV.

For HD\,113703 we show a somewhat larger image 
to include the detection of the wide Lindroos companion HD113703\,B.
Adopting the {\em Hipparcos} position for HD\,113703\,B 
(position angle, PA, of $73^\circ$, separation of $11.2^{\prime\prime}$) 
the distance between this star and the X-ray source is quite large ($1.2^{\prime\prime}$). 
We have examined our optical images of this star taken in February 2001 by one of us (NH), 
and find HD\,113703\,B at a position angle of $79^\circ$ and separation of 
$11.5^{\prime\prime}$ from the B-type star. Comparing the position of HD\,113703\,B
in our optical images with the position of the X-ray source we find that the displacement
is only $0.2^{\prime\prime}$. Therefore, we suppose that the PA cited in the 
{\em Hipparcos} data base is erroneous. 

In Table~\ref{tab:params} we list the X-ray parameters for the X-ray sources
shown in Fig.~\ref{fig:acis_images}.
We give HD number (Col.~1), X-ray position (Cols.~2 and~3),
separation to B-type star and new companion 
(Cols.~4 and~5), significance of detection
(Col.~6), exposure time (Col.~7). The total number of counts (Col.~8) and
the X-ray luminosity (Col~10) refer to the $0.5-8$\,keV passband.
$L_{\rm x}$ was derived by integrating the ACIS spectrum 
(see Sect.~\ref{sect:spectra} and Sect.~\ref{sect:lx}). 
It was assumed that the systems are physically bound, and all components are
located at the distance of the B-star given in Table~\ref{tab:obslog}. 
This may not be true in all cases (see Sect.~\ref{sect:cmd}).
Col.~9 is the hardness ratio defined as 
$HR = (H-S)/(H+S)$, where $H$ and $S$ are the number of counts in a hard band
($1-8$\,keV) and soft band ($0.5-1$\,keV), respectively. 
The last column is the X-ray temperature of a isothermal model fit to the
ACIS spectrum (see Sect.~\ref{sect:spectra} for details).

%
% * X-ray positions by converting RA and DEC of wavdetect source list
%   using get_srcinfo.pro
% * 
% Exposure times from fits header of final events file
%
\begin{table*}
\begin{center}
\caption{\small  Positions and X-ray parameters of X-ray sources associated with late-B type stars or their IR companions at separations between $1-6^{\prime\prime}$.}
\label{tab:params}
\small
\begin{tabular}{lrrrrrrcrrr} \hline
HD & \multicolumn{1}{c}{$\alpha_{\rm x,2000}$} & \multicolumn{1}{c}{$\delta_{\rm x, 2000}$} & Sep X-A                                & Sep X-B                                & Sign. & \multicolumn{1}{c}{Expo} & ACIS       & \multicolumn{1}{c}{HR} & $\lg{L_{\rm x}}^*$ & $kT$ \\ 
   &                                           &                                            & \multicolumn{1}{c}{[${\prime\prime}$]} & \multicolumn{1}{c}{[${\prime\prime}$]} &       & \multicolumn{1}{c}{[s]}  & counts$^*$ &                        & [erg/s] & [K] \\ 
\hline
\multicolumn{10}{c}{\bf Obs-ID 2541, Seq.No 200149} \\
1685\,X-ray\,1   & 00 20 38.99 & $-$69 37 29.70 & 0.07 & 2.25 & 21.5 & 2338 & $44.0 \pm 6.6$ & $-0.09 \pm 0.21$ & $29.1$ & $0.53$ \\
1685\,X-ray\,2   & 00 20 38.75 & $-$69 37 31.66 & 2.35 & 0.11 & 33.8 & 2338 & $71.0 \pm 8.4$ & $0.07 \pm 0.17$  & $29.4$ & $1.06$ \\
\hline
\multicolumn{10}{c}{\bf Obs-ID 0626, Seq.No 200051} \\
113703\,X-ray    & 13 06 16.57 & $-$48 27 47.97 & $1.37$ & $0.21$ & $299.5$ & 12184 & $1327.0 \pm 36.4$ & $-0.35 \pm 0.04$ & $29.8$ & $0.76$ \\
\hline
\multicolumn{10}{c}{\bf Obs-ID 2542, Seq.No 200150} \\
123445\,X-ray    & 14 08 52.35 & $-$43 28 12.57 & 5.39 & 0.15$^\dagger$ & 33.3 & 2237 & $65.0 \pm 8.1$ & $0.05 \pm 0.18$  & $29.8$ & $0.87$ \\
\hline
\multicolumn{10}{c}{\bf Obs-ID 2543, Seq.No 200151} \\
133880\,X-ray    & 15 08 12.24 & $-$40 35 02.49 & 1.66 & 0.46 & 88.6 & 2461 & $218.0 \pm 14.8$ & $0.17 \pm 0.09$ & $30.0$ & $1.05$ \\
\hline
\multicolumn{10}{c}{\bf Obs-ID 2544, Seq.No 200152} \\
169978\,X-ray    & 18 31 22.42 & $-$62 16 42.01 & 0.20 & 2.92 & 57.1 & 2420 &$125.0 \pm 11.2$ & $0.12 \pm 0.13$ & $29.9$ & $0.83$ \\
\hline
\multicolumn{9}{l}{$^*$ in the $0.5-8$\,keV passband; $L_{\rm x}$ refers to the distance given in Table~\ref{tab:obslog}.} \\
\multicolumn{9}{l}{$^\dagger$ This is the distance to IR-companion 'D'. The distance to IR-companion 'C' is $0.17^{\prime\prime}$.}
\end{tabular}
\end{center}
\end{table*}

\section{Spectral Analysis}\label{sect:spectra}

The high sensitivity of ACIS jointly with the improved spectral resolution
as compared to the {\em ROSAT} PSPC allows for the first time to examine the
characteristics of these stars by means of a direct spectral
analysis. 
Except for Obs-ID\,$0626$ the number of counts collected per X-ray source 
are rather small due to the low
exposure time, but sufficient for a basic description of the temperature
of the emitting plasma.

For each of the X-ray sources listed in Table~\ref{tab:params} we
extracted a spectrum, the corresponding detector response matrix that maps
pulse heights into energy space, and an
auxiliary response file which contains information about the effective area
and detector efficiency across the chip as a function of energy.
We binned each spectrum to a minimum of 10~counts per bin. 
As the background of ACIS is very low 
(measured to be $< 1$\,count in the source extraction area)
it can be neglected. 
Spectral modelling was performed in the XSPEC environment, version 11.2.0. 

Version 2.3 of the CIAO tools is the first one that implements a correction for 
the charge transfer inefficiency (CTI). 
CTI affects the spectrum of astrophysical sources by introducing an apparent 
gain shift and degrading the energy resolution. 
Currently the CTI correction is not included in the pipeline processing
performed at the {\em Chandra} X-ray Center (CXC).
Therefore, to apply this correction to the data the level\,1 events file 
has to be reprocessed by the user and converted to a level\,2 events file. 
After taking account of the CTI correction we encountered serious
problems in the spectral fitting process. None of the fits converged to an
acceptable solution. Therefore, we decided to work on the uncorrected spectrum,
until more reliable tools allow to consider the CTI effect. In any case, 
due to the low number of counts in our spectra this effect should not exceed 
the statistical uncertainties. 

\begin{figure*}
\begin{center}
\parbox{18cm}{
\parbox{9cm}{
\resizebox{9cm}{!}{\includegraphics{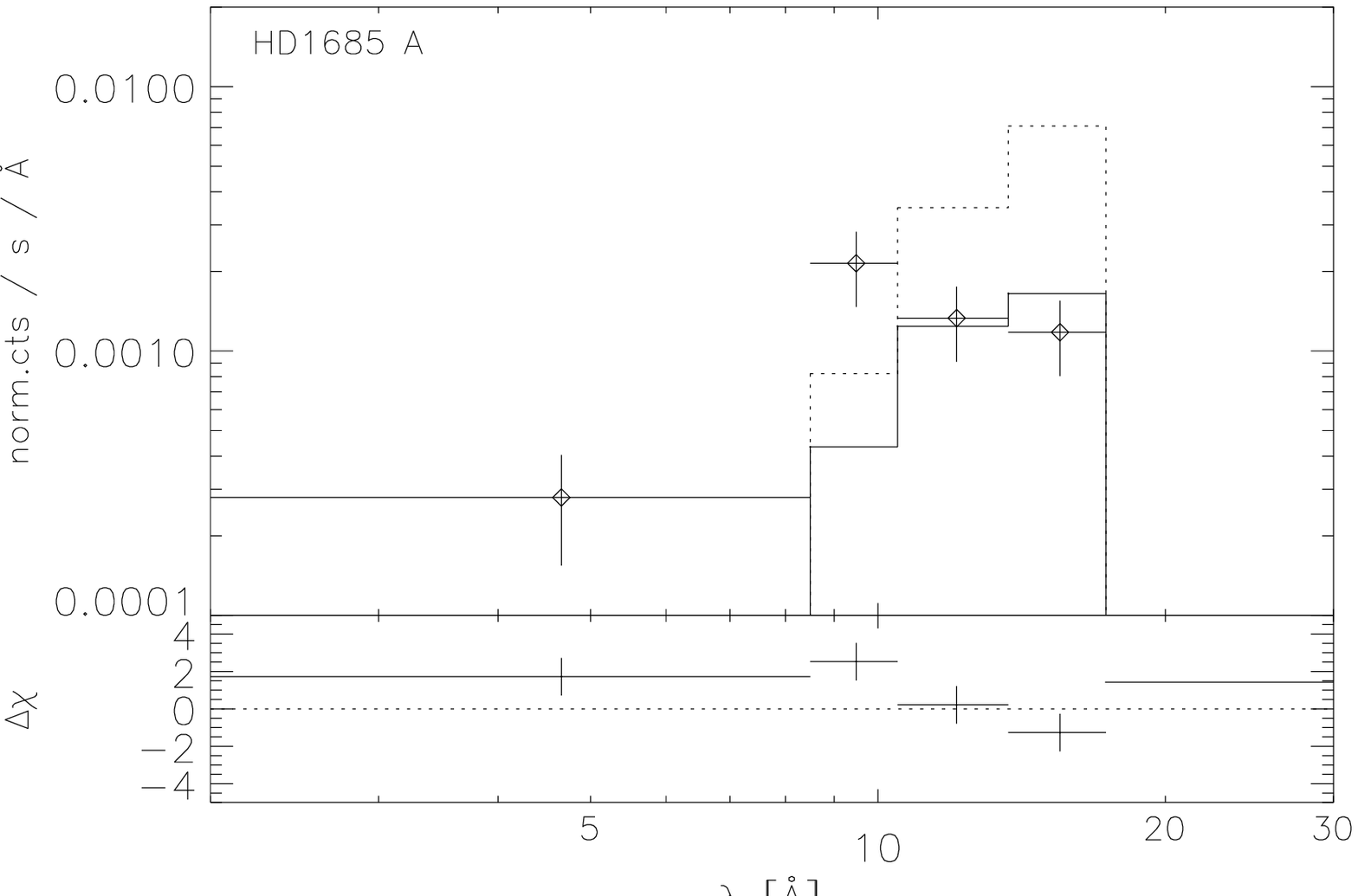}}
}
\parbox{9cm}{
\resizebox{9cm}{!}{\includegraphics{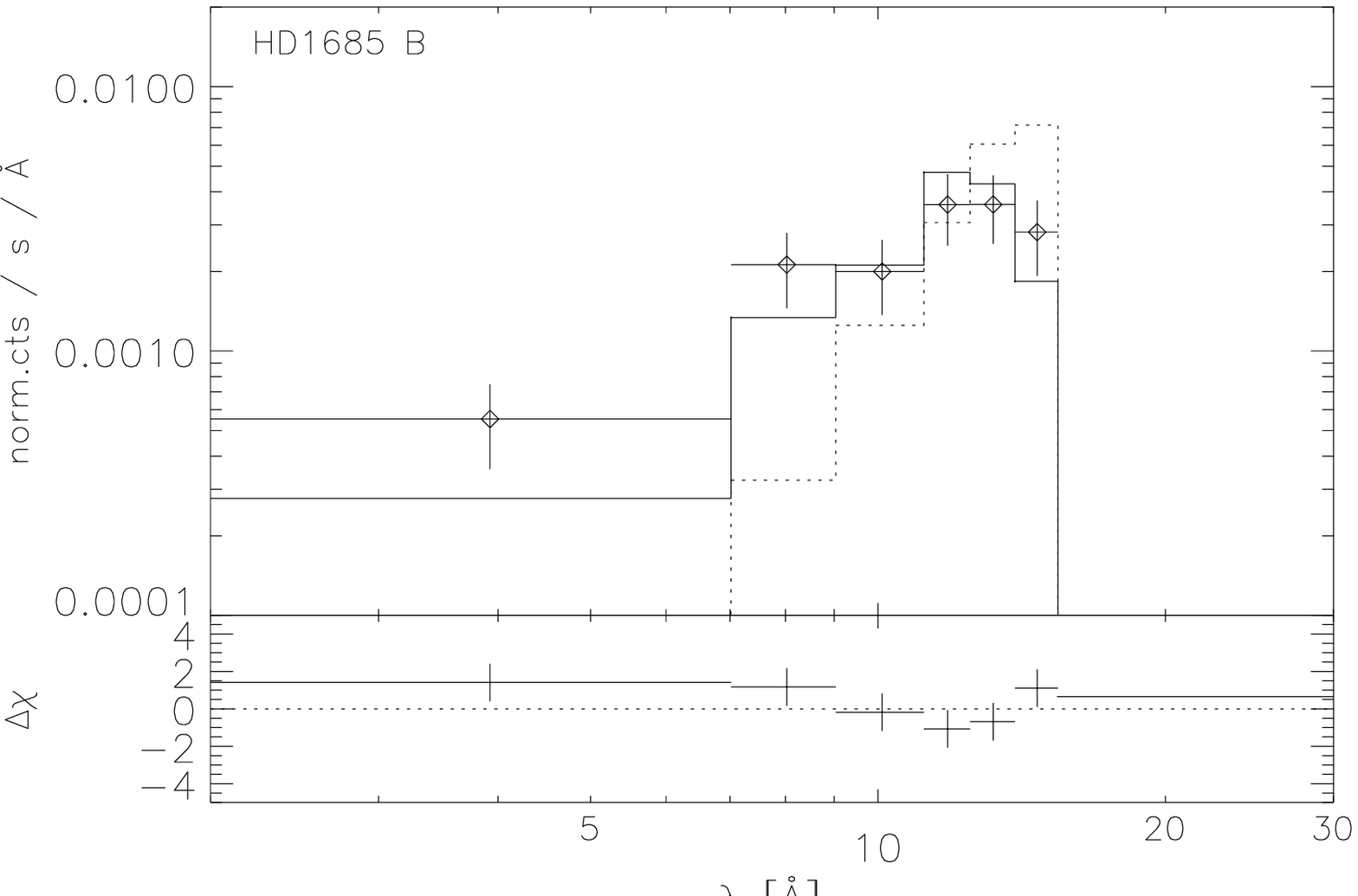}}
}
}
\parbox{18cm}{
\parbox{9cm}{
\resizebox{9cm}{!}{\includegraphics{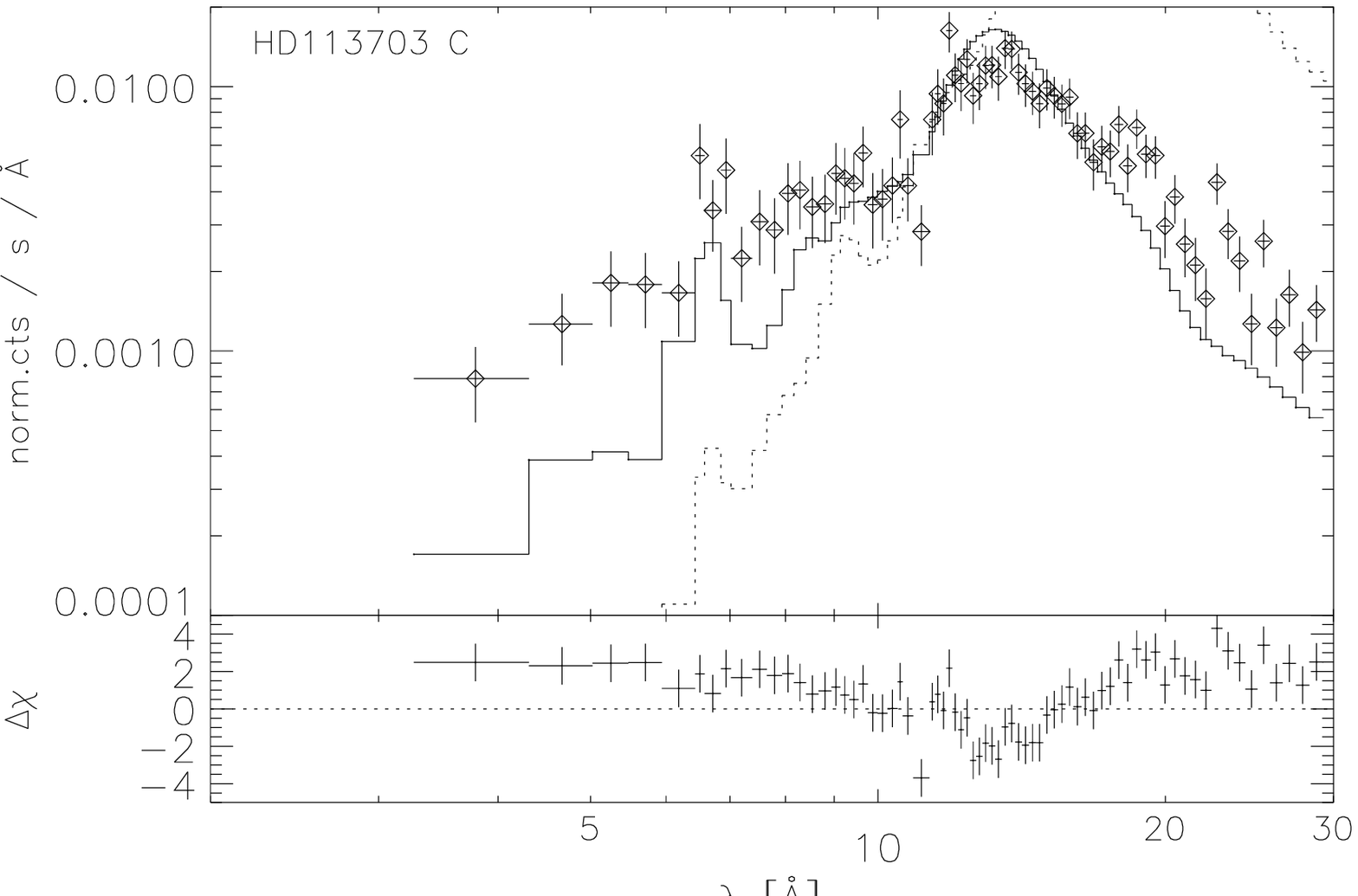}}
}
\parbox{9cm}{
\resizebox{9cm}{!}{\includegraphics{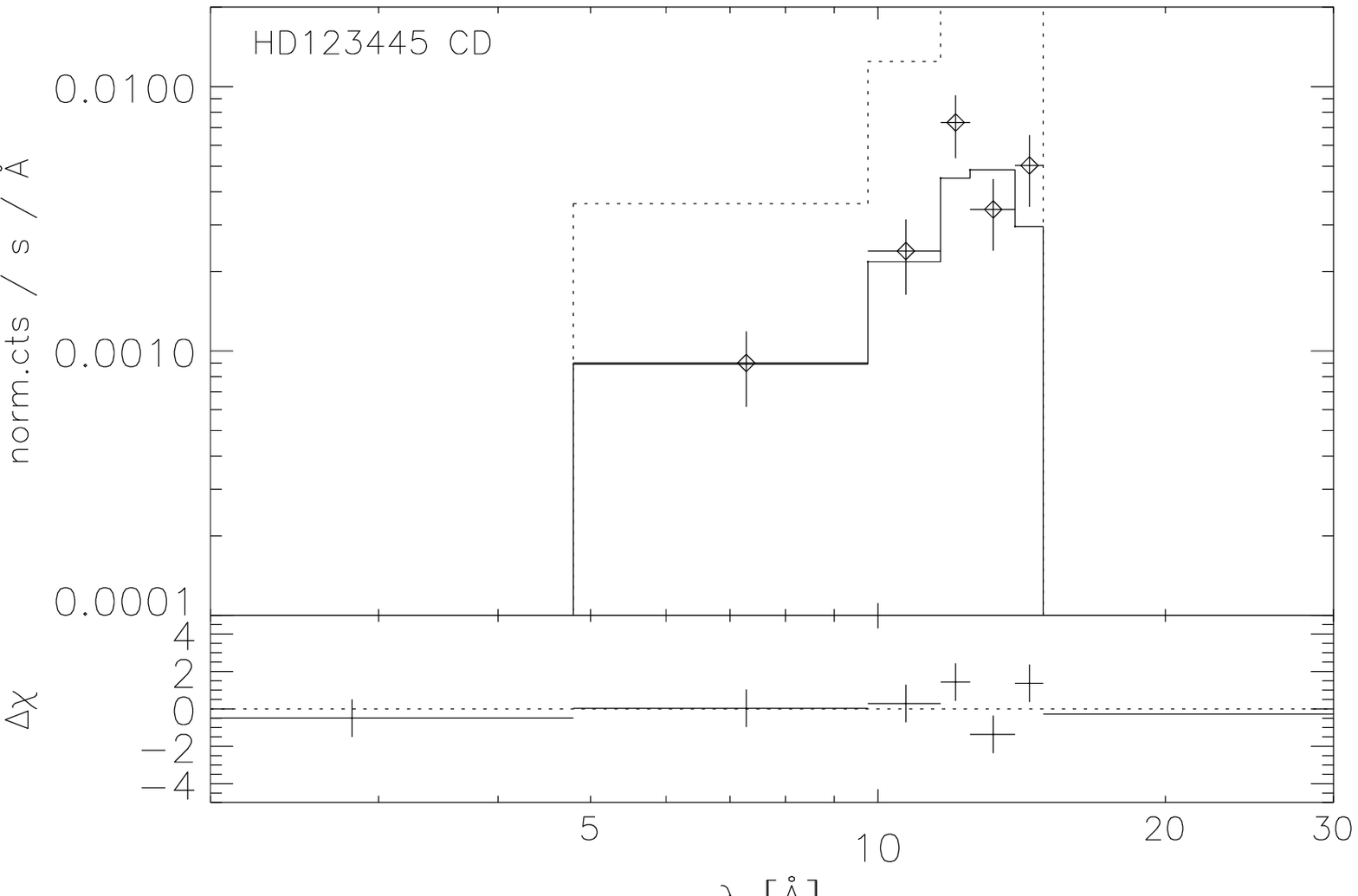}}
}
}
\parbox{18cm}{
\parbox{9cm}{
\resizebox{9cm}{!}{\includegraphics{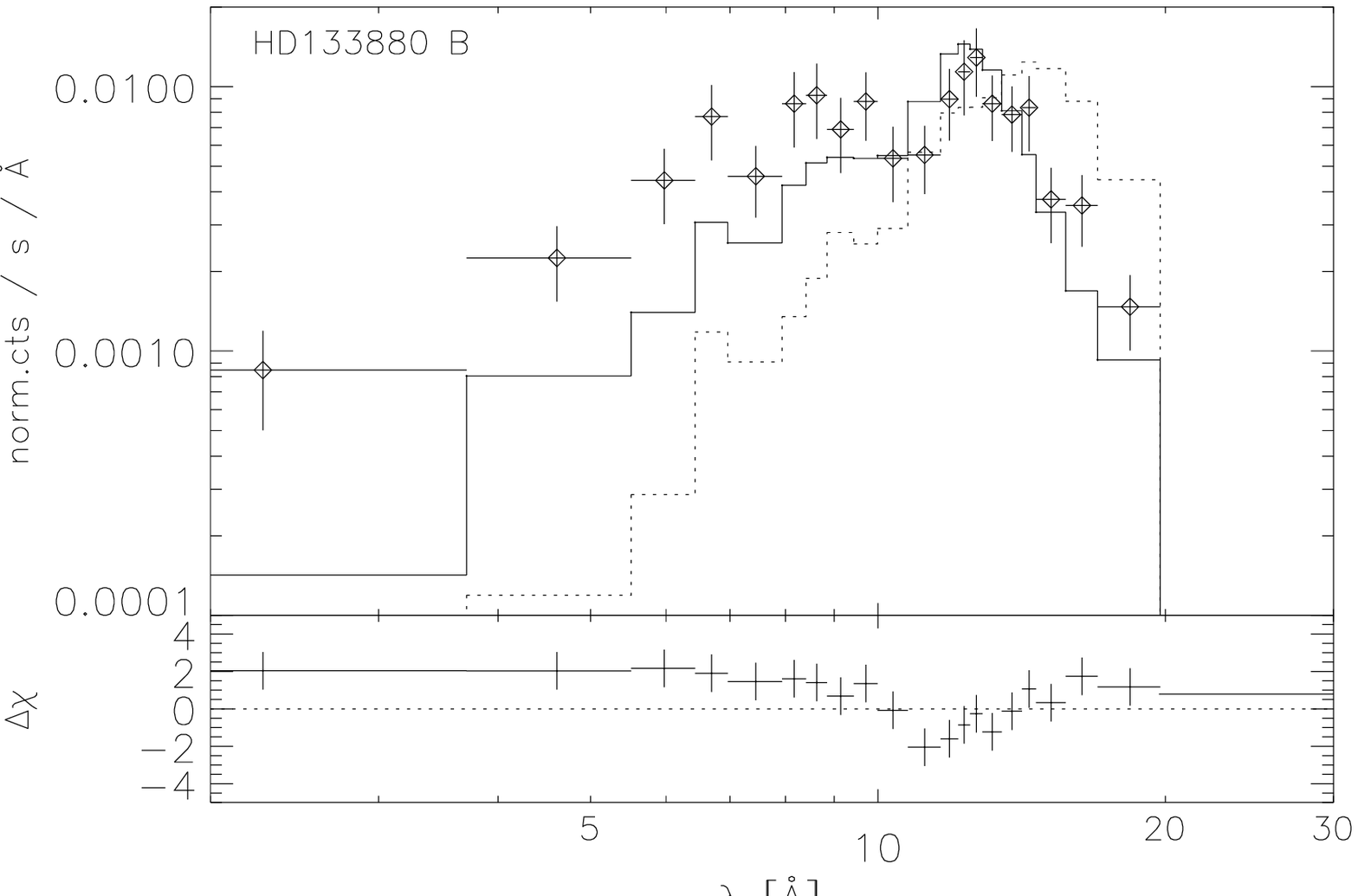}}
}
\parbox{9cm}{
\resizebox{9cm}{!}{\includegraphics{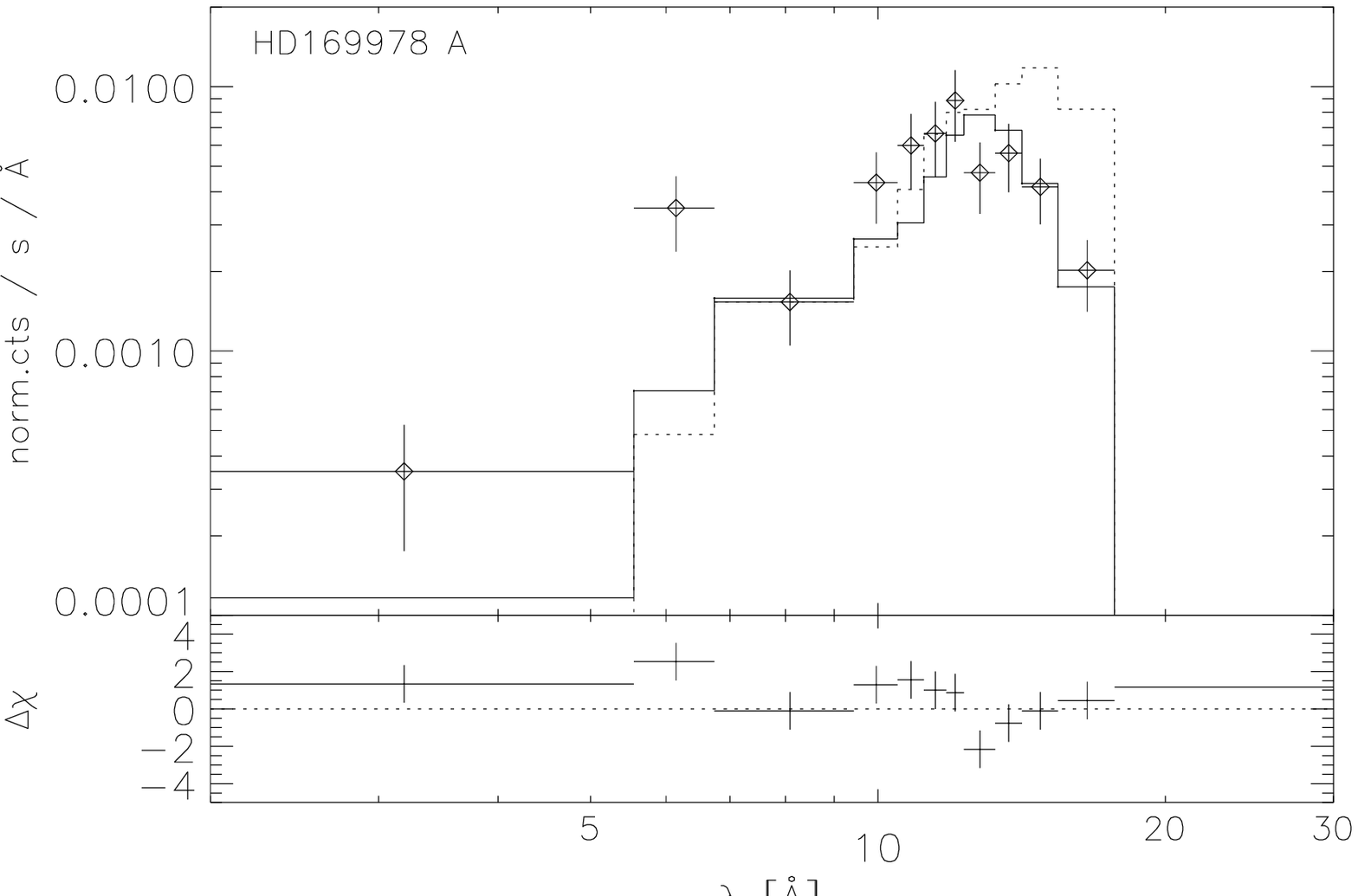}}
}
}
\caption{ACIS spectra of X-ray sources identified with a component of the B-star systems. {\em solid lines} - best fit of 1-T model with solar abundances (see text in Sect.~\ref{sect:spectra} and Table~\ref{tab:params}), {\em dotted lines} - simulated spectrum obtained for spectral parameters estimated from {\em ROSAT} observations (\protect\cite{Berghoefer96.1} and \protect\cite{Huelamo00.1}).}
\label{fig:acis_spec}
\end{center}
\end{figure*}

First we approximated each spectrum 
with a one-temperature (1-T) thermal model (MEKAL) including a 
photo-absorption term comprising the atomic cross-section by 
\citey{Morrison83.1}. 
For three of the systems the optical extinction is known (see \cite{Lindroos86.1}),
and $N_{\rm H}$ can be estimated using the relation by \citey{Paresce84.1} to be
$\leq 2\,10^{20}\,{\rm cm^{-2}}$. The other two are likely to have similarly low
values of absorption because of their proximity which provides low interstellar
column density, and because at their age of $\leq\,100$\,Myrs no substantial 
amount of circumstellar material is present. 
As the spectral fits showed a tendency to converge towards much higher values
for $N_{\rm H}$ we imposed the constraint that the column density may not
exceed $8\,10^{20}\,{\rm cm^{-2}}$. 
The temperatures derived from the 1-T fits with solar abundances are presented
in Table~\ref{tab:params}. In Fig.~\ref{fig:acis_spec} we overlay the data
by this model.  

Noticeably this model is not adequate for most of the stars. Therefore,  
we refined our approach and allowed for variable abundances. This results in an 
acceptable solution for HD\,1685\,X-2 with $Z \sim 0.16\,Z_\odot$ 
($\chi^2_{\rm red} = 0.9$ for $3$ dofs), and presents a slight improvement in
$\chi^2$-statistics for 
HD\,133880 ($Z \sim 0.18\,Z_\odot$ and $\chi^2_{\rm red} = 1.4$ for $17$ dofs)
and 
HD\,169978 ($Z \sim 0.20\,Z_\odot$ and $\chi^2_{\rm red} = 1.9$ for $8$ dofs). 
However, in the latter two cases substantial residuals remain.
The clearest improvement is found for HD\,113703, where
$Z \sim 0.15\,Z_\odot$ and $\chi^2_{\rm red} = 1.4$ for $17$ dofs.
For HD\,1685\,X-1 and HD\,123445 
the poor quality of the spectrum does not justify the use of a more refined model.

To remove the remaining residuals in the spectrum of 
HD\,113703\,C, HD\,133880\,B, and HD\,169978\,A 
introducing a second temperature is in order. 
We find hotter components in the range $1.5-3.5$\,keV, and cooler components 
only slightly smaller than in the 1-T models. 
However, due to the low statistics at high energies the hotter temperatures 
are poorly constrained. Indeed, the parameter space of the 2-T model shows
several local minima depending on the initial values of the temperatures. 
We mention in passing that 
the elevated emission near $6$\,\AA~ in the spectrum of HD\,169978\,A  
is likely to be explained by enhanced silicon abundance. But
the low number of counts prohibits the determination of the abundances of
individual elements. 

We stress that despite the short exposure times and the problems outlined
above with the 2-T fits 
the derived parameters are not meaningless. In particular, does the 
1-T model reproduce the temperature of the bulk of the emitting material, although
the detailed temperature structure of the corona is inaccessible.  
\citey{Feigelson02.1} have shown for similar models fitted
to X-ray faint young stars in Orion that the typical uncertainties in $kT$ 
for sources between $30$ and $100$ counts with ACIS are about $30...60$\,\%. 
Thus, our {\em Chandra} observations allow for the first time a quantitative
assessment of the X-ray temperature of these stars. 
An earlier study of the X-ray emission from early-type stars 
by \citey{Berghoefer96.1} based on {\em ROSAT} All-Sky Survey (RASS) 
observations used the
{\em ROSAT} PSPC hardness ratio in comparison with a grid of model spectra 
generated for various column densities and X-ray temperatures 
to estimate the conditions in the emitting region. 
They concluded that most OB stars have a very soft X-ray spectrum,
with a typical temperature $T_{\rm x} < 0.5$\,keV. This is in clear contrast to
our {\em Chandra} observations: All our targets require temperatures in excess
of $0.5$\,keV, and the three stars with the highest S/N among our targets 
even suggest a second spectral component with temperature above $1.5$\,keV.  

The discrepancies between the {\em ROSAT} estimate and the {\em Chandra} spectra 
are graphically demonstrated in Fig.~\ref{fig:acis_spec}: The dotted lines are
simulated ACIS spectra based on the spectral parameters listed by 
\citey{Berghoefer96.1}. Only HD\,123445 was not detected in the RASS, and we use
the parameters given by \citey{Huelamo00.1} extracted from a {\em ROSAT} HRI
observation. 
The simulations were performed with XSPEC for a 1-T model 
with $N_{\rm H}$ and $kT$ fixed on the {\em ROSAT} values, and the normalization
was varied until the X-ray luminosities given by \citey{Berghoefer96.1}
and \citey{Huelamo00.1} were reproduced. 
We also took into account that \citey{Berghoefer96.1} have used their own
estimate of photometric distances, which are different from the ones in 
Table~\ref{tab:obslog}. Fig.~\ref{fig:acis_spec} shows clearly that {\em ROSAT}
has misestimated both the brightness and spectral distribution of these
sources, presumably due to the low S/N of the data which necessitated 
(wrong) model assumptions.

\section{X-ray Luminosities}\label{sect:lx}

The ACIS observations at hand facilitate the derivation of X-ray luminosities 
of the detected sources directly by integrating the spectrum. The results are
listed in Table~\ref{tab:params} for the $0.5-8$\,keV energy band. 
We add here some remarks on the individual objects and present upper limits
for the undetected components. 

The {\em Chandra} observation of HD\,1685 has shown that {\em ROSAT} had 
confused two X-ray sources of similar brightness, and that the X-ray luminosity 
of the ACIS source at the position of HD\,1685\,B is lower than thought before. 
Both the X-ray temperature and luminosity of HD\,1685\,A are similar to the detected 
late-type stars, and therefore point at the presence of 
another as yet undiscovered late-type companion. 

For HD\,113703 \citey{Berghoefer96.1} provided a value for $L_{\rm x}$ which is
too high by one order of magnitude. This difference can not be explained by the
smaller energy range of the {\em ROSAT} PSPC.
In principle it could be due to variability of the star, but a mis-estimate
of the {\em ROSAT} luminosity resulting from the method used by \citey{Berghoefer96.1}
(see Sect.~\ref{sect:spectra}) seems more probable. 

The possible nature of HD\,123445\,CD was 
discussed by \citey{Huelamo01.1}. A binary composed of two K-type stars was
put forth as most likely because the {\em ROSAT} luminosity and hardness were
not in agreement with them being foreground M-type stars or background giants.
\citey{Huelamo01.1} argued that if HD\,123445\,CD were indeed K-type stars 
their IR colors indicate a distance of $\sim 140$\,pc (consistent with the mean
distance of the Upper-Centaurus-Lupus (UCL) association; \cite{deZeeuw99.1}).  
In this case they may not form a physical pair with HD\,123445\,A.
Accordingly their X-ray luminosity would be lower than given in 
Table~\ref{tab:params}, namely $\lg{L_{\rm x}} = 29.4$\,erg/s.
Regardless on whether the distance is $218$ or $140$\,pc the {\em ROSAT}
estimate for $L_{\rm x}$ was far too high. As in the case of the estimates
by \citey{Berghoefer96.1} discussed above this is presumably a result of the
assumptions on the spectral properties of the source   
made by \citey{Huelamo00.1}. 

\citey{Hubrig01.1} noticed that the {\em ROSAT} X-ray luminosity of HD\,169978 
is too high in terms of $\lg{(L_{\rm x}/L_{\rm bol})}$ 
to be emitted from the companion, which is a very-low mass star 
($M_{\rm B}=0.15\,M_\odot$).
Our {\em Chandra} observation has assigned the X-ray source to the B-type star
which is meanwhile known to be a spectroscopic binary. The X-ray luminosity 
of $8\,10^{29}$\,erg/s provided by ACIS at the position of HD\,169978\,A is  
consistent with the earlier {\em ROSAT} measurement.

To compute upper limits for the undetected components of our target systems 
we used the method described by \citey{Kraft91.1}. 
We count the photons in a circle of $1.25^{\prime\prime}$ radius centered on the
optical/IR position of the star. This photon extraction area is similar to the
{\em wavdetect} source ellipses of the detected sources. 
For HD\,133880\,A and HD\,113703\,A these circles overlap with those
of their nearby detected companions. Therefore, we use the number of photons
collected in the semicircle pointing away from the companion, and extrapolate
this value to the full extraction area. 
Not a single photon was collected at the position of HD\,123445\,A.
We measure 6~counts at the IR position of HD\,169978\,B.
We took account of the background 
fluctuations by estimating the background within a squared area of 
$1^\prime$ radius centered on the optical/IR position of the respective
star, and scaling this mean background to 
the source extraction area. 
Using these numbers in connection with the values tabulated by 
\citey{Gehrels86.1} provides the number of upper limit counts. The conversion to 
X-ray flux and luminosity is performed with help of 
PIMMS\footnote{http://asc.harvard.edu/toolkjet/pimms.jsp}
assuming a 10\,MK hot thermal plasma and negligible absorption. 
We tabulate the resulting X-ray luminosities for
the $0.5-8.0$\,keV band in Table~\ref{tab:upperlimits}.
\begin{table}
\begin{center}
\caption{95\,\% confidence upper limits to the X-ray luminosity of 
undetected primaries and new IR objects; numbers are for the ACIS broad band 
($0.5-8$\,keV).}
\label{tab:upperlimits}
\begin{tabular}{lrrrr} \\ 
\noalign{\smallskip}\hline\noalign{\smallskip}
                          & 113703\,A & 123445\,A & 133880\,A & 169978\,B \\
\noalign{\smallskip}\hline\noalign{\smallskip}
$\lg{L_{\rm x}}$\,[erg/s] & $<27.85$  & $<28.54$  & $<28.02$  & $<28.76$ \\
\noalign{\smallskip}\hline\noalign{\smallskip}
\end{tabular}
\end{center}
\end{table}

\subsection{The $L_{\rm x}/L_{\rm bol} - $relation}\label{subsect:lxlbol}

The ratio between X-ray and bolometric luminosity, 
$L_{\rm x}/L_{\rm bol}$, is a crucial indicator for stellar activity. 
The most active late-type stars 
-- generally coinciding with the most rapid rotators -- 
are observed to display values near $10^{-3}$.
An unidentified mechanism seems to prevent the generation of X-rays above 
this limit. Several possible causes for the saturation phenomenon are discussed,
such as stripping of the corona by centrifugal forces (\cite{Jardine99.1})
or complete filling of the stellar surface with active regions (\cite{Vilhu84.1}). 
Less active late-type stars range between $L_{\rm x}/L_{\rm bol} = 10^{-4...-5}$. 
The spread is thought to be caused by the influences of various stellar parameters
such as mass, rotation, and age on the level of X-ray emission. 
Hot stars, for which X-ray emission is thought to arise in a stellar wind,
are clearly distinct from late-type stars with a typical value of 
$L_{\rm x}/L_{\rm bol} \approx 10^{-7}$.

\begin{figure}
\begin{center}
\resizebox{9cm}{!}{\includegraphics{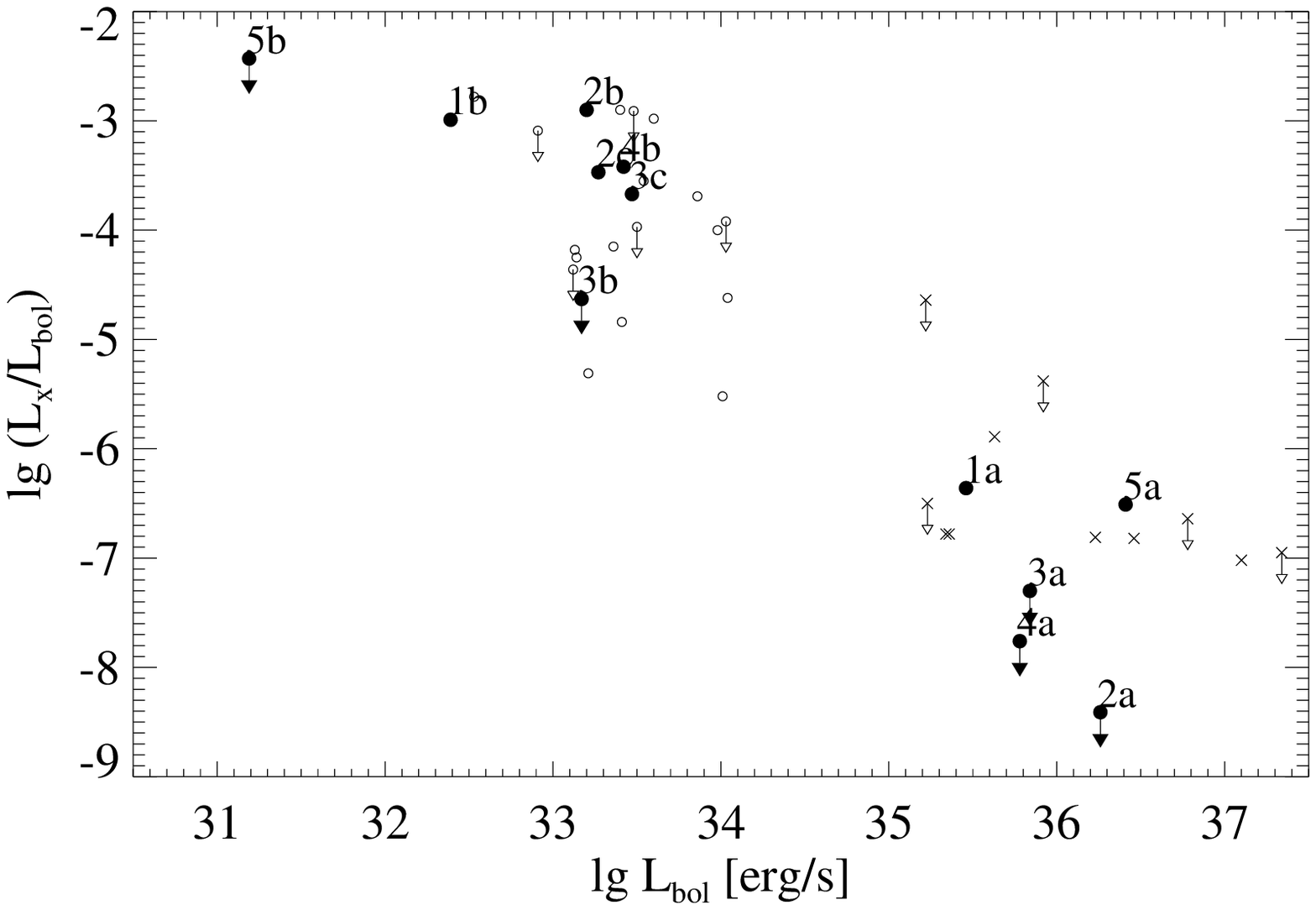}}
\caption{Ratio between X-ray and bolometric luminosity $\lg{(\rm L_{\rm x}/L_{\rm bol})}$ versus bolometric luminosity. {\em filled circles} - detections and upper limits derived in this paper; individual objects: [1] - HD\,1685, [2] - HD\,113703, [3] - HD\,123445, [4] - HD\,133880, [5] - HD\,169978; for Lindroos systems [b] denotes the Lindroos secondary and [c] the new IR object, else [b] is the IR object. Note that if the 'secondaries' were not bound to the B-type stars their distance would be different from the value given in Table~\ref{tab:obslog}, and our estimate of the luminosities would be wrong. 
{\em open circles} - late-type stars in Lindroos systems observed with {\em ROSAT} (see \protect\cite{Huelamo00.1}), {\em x-points} - early-type stars in Lindroos systems observed with {\em ROSAT} (see \protect\cite{Huelamo00.1}).}
\label{fig:acis_lx_lbol}
\end{center}
\end{figure}

Fig.~\ref{fig:acis_lx_lbol} displays the $L_{\rm x}/L_{\rm bol}$ ratio 
for all components of the B-star systems observed with {\em Chandra}. 
The bolometric luminosities for the low-mass components (both the Lindroos
secondaries and the new companions) are given in Table~\ref{tab:cmd} 
and~\ref{tab:lindroos} of Sect.~\ref{sect:cmd} and 
Sect.~\ref{sect:lindroos}, where we explain also how they were derived. 
For comparison, in Fig.~\ref{fig:acis_lx_lbol} 
we show also the Lindroos systems observed with {\em ROSAT} (\cite{Huelamo00.1}). 

In the sample
investigated here all new IR objects display $\lg{(L_{\rm x}/L_{\rm bol})}$ 
values near the saturation limit. For HD\,169978\,B only an upper limit
is derived, and due to its small
bolometric luminosity the $\lg{(L_{\rm x}/L_{\rm bol})}$ ratio is still
rather ill-constrained, although the sensitivity is improved by more than one 
order of magnitude with respect to previous X-ray measurements for this 
stellar system. 
For the three undetected B-type stars the upper limits to 
$\lg{(L_{\rm x}/L_{\rm bol})}$ are below $10^{-7}$.
The two detected early-type stars show $\lg{(L_{\rm x}/L_{\rm bol})} \sim -6.5$, 
similar to that of most of the primaries of the 
Lindroos sample studied with {\em ROSAT}. 

\section{Evolutionary state of the system components}\label{sect:cmd}

Next we examine whether the low-mass companions and the B-type primaries
are physical rather than optical systems 
by comparing their evolutionary stage. 
If they form bound systems, both the late B-type star and the IR object 
should have the same age.

\subsection{Late B-type primaries}\label{subsect:cmd_prim}

The primaries in the stellar systems under consideration have been 
compared to evolutionary models by 
\citey{Hubrig01.1} and \citey{Gerbaldi01.1}, who  
found that their ages range from $\sim 50...200$\,Myrs.  
On the other hand three of the targets are known to be members of different 
OB~associations: HD\,113703\,A is believed to belong to the Lower Centaurus
Crux subgroup of the ScoCen OB\,2 association, while HD\,123445\,A and
HD\,133880\,A are members of the UCL co-moving group
(\cite{deZeeuw99.1}). The Sco-Cen OB\,2 subgroups have ages between 
$5-13$\,Myrs (\cite{deGeus89.1}), about one order of magnitude lower
than the individual ages derived by \citey{Hubrig01.1} and 
\citey{Gerbaldi01.1} from model tracks. 
We suppose that this discrepancy arises from 
uncertainties in the stellar parameters 
that can be summarized as follows: 
\begin{itemize}
\item For HD\,133880\,A, e.g., the $B-V$ color varies
with an amplitude of $\sim 0.1$\,mag, leading to a spectral type range of
B3...B9 according to \citey{Kenyon95.1}, and an uncertainty of at least $\pm 2000$\,K 
in $T_{\rm eff}$ allowing for an age as small as $\sim 10$\,Myrs. 
\item For HD\,123445\,A a major uncertainty
stems from the distance. This star has a {\em Hipparcos} parallax of 
$218 \pm 40$\,pc, while the distance to the UCL association is $140$\,pc
(\cite{deZeeuw99.1}). Thus $L_{\rm bol}$ is not constrained very well,
and the age of HD\,123445\,A may be smaller than suggested by 
\citey{Gerbaldi01.1}.  
\item HD\,169978\,A is listed as a possible member
of the Wolf\,630 moving group (\cite{McDonald83.1}). The age of the Wolf\,630  
cluster is $\sim 5$\,Gyrs. Therefore, HD\,169978\,A could already
have left the MS, consistent with its luminosity class (III) found in
SIMBAD. However, the membership of HD\,169978\,A to Wolf\,630 
has not been established firmly. Rather its bolometric luminosity and effective temperature
place it near the end of the MS, and taking the newly discovered
spectroscopic companion into account 
should move its position further down in the HR diagram.
\end{itemize}

\subsection{IR secondaries}\label{subsect:cmd_seco}

In order to examine the 
evolutionary state of the ADONIS companions we compare their IR  
magnitudes and colors with models for low-mass pre-MS stars. 
We use the calculations by \citey{Baraffe98.1}
with helium abundance of $Y = 0.275$, $[M/H] = 0$, and mixing length parameter 
of one pressure scale height ($\alpha_{\rm ML} = 1$). Models with $[M/H] = -0.5$, $Y = 0.25$, and 
$\alpha_{\rm ML} = 1.9$ are also available, as well as models with $[M/H] = 0$ and $Y = 0.282$. 
The former one reproduces the present-day Sun but has been calculated only for
a small range of masses. 

\begin{figure}
\begin{center}
%\resizebox{9cm}{!}{\includegraphics{/home/stelzer/data/publ/bstars/ps/baraffe98_cmd_hd123445-218.eps}}
%\resizebox{9cm}{!}{\includegraphics{/home/stelzer/data/publ/bstars/ps/baraffe98_cmd_hd123445-140.eps}}
\resizebox{9cm}{!}{\includegraphics{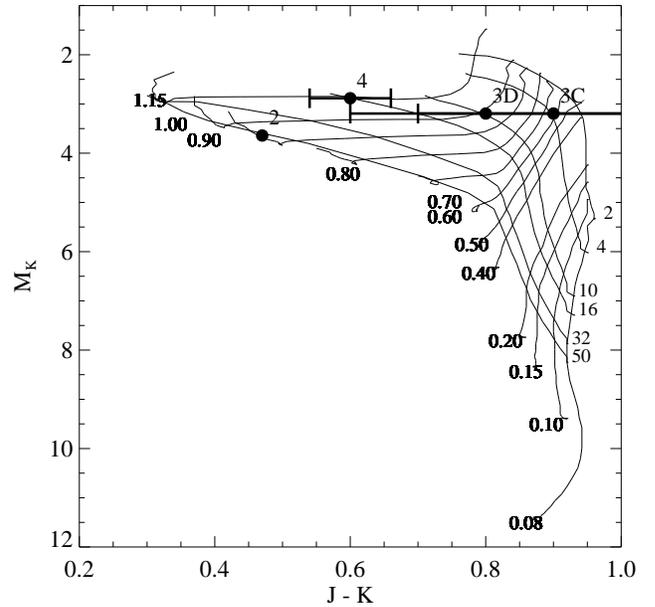}}
\caption{Infrared color-magnitude diagram according to the evolutionary models by \protect\citey{Baraffe98.1} for $Y=0.275$ and $\alpha_{\rm ML}=H_{\rm p}$. Filled circles indicate the positions of new companions: [2] - HD\,113703, [3] - HD\,123445\,C and~D, [4] - HD\,133880. The IR magnitudes of the companion to HD\,1685 do not allow to place it onto the tracks, and for the companion to HD\,169978 no $J$-band image is available.}
\label{fig:cmd}
\end{center}
\end{figure}

In Fig.~\ref{fig:cmd} we show the Baraffe model in the $M_{\rm K}$
versus $J-K$ diagram. To position the ADONIS objects 
on the tracks and isochrones we made use
of published $J$ and $K$ magnitudes (see references in 
Table~\ref{tab:cmd}), and we assumed
that all companions are bound to their primaries, i.e. their distances are those
listed in Table~\ref{tab:obslog}. 
Estimates for mass, age, luminosity, and effective temperature 
for those companions that are compatible with the pre-MS evolutionary tracks 
are obtained by interpolating the tracks,
and are given in Table~\ref{tab:cmd}. 
Next we discuss the individual objects: 
\begin{itemize} 
\item For the companion to HD\,169978 only a $K$ band image was taken, and we can not 
put it into the color-magnitude diagram (CMD). 
Therefore its evolutionary stage remains unclear. 
\item As pointed out by \citey{Hubrig01.1} the companion of HD\,1685
has near-IR colors inconsistent with pre-MS models, 
questioning that this object is physically bound to the B-type star. 
\item The error in the $J-K$ color of the companions to HD\,123445 are large. 
Because of this and the large uncertainty of the distance of HD\,123445 
the mass and age
of HD\,123445\,C and~D are not well constrained. 
But if assumed to be at $218$\,pc both objects are most likely younger than $30$\,Myrs. 
Therefore we conclude that being a member of
the young UCL association (age $\sim 10$\,Myrs; \cite{deGeus89.1}) 
is favored for both new IR sources near HD\,123445. 
While uncompatible with the age of $160$\,Myrs that \citey{Gerbaldi01.1} derived for 
HD\,123445\,A, it seems quite possible that they are physically bound to the B-type star,
if the actual age of the latter one is smaller as discussed above. 

In May 2001, i.e. about one year after the first ADONIS images were taken, 
the AO observations of HD\,123445\,CD were repeated. 
The aim of this effort was to measure relative movements between the components C and D
to track down their orbital motion, or their proper motion with respect to A. 
In the new IR images we measure nearly the same separation and position angle 
of~C and~D with respect to the primary as before (Hu\'elamo et al., in prep.). 
The fact that both components
have not moved contrasts with the previous suspicion that they may be unrelated
foreground M-type stars. 
But we still can not exclude the possibility that the two objects are 
K-type stars at $140$\,pc. For this distance  
the errors in the IR photometry allow for HD\,123445\,C and~D a somewhat older age,
but are still compatible with them being on the pre-MS. So the conclusion that these
stars are UCL members remains unchanged. 

\item For HD\,113703\,C the
error in the IR color is negligible, and thus the mass and age are much better defined.
Furthermore, the {\em Hipparcos} distance for HD\,113703 is in agreement with the mean
distance of the LCC association to which this star belongs according to \citey{deZeeuw99.1}.
We derive an age of $50$\,Myrs for HD\,113703\,C consistent with the age of the B-type star
($48$\,Myrs according to \cite{Gerbaldi01.1}). 
This suggests that HD\,113703\,C is a true companion to HD\,113703\,A. 

\item According to Fig.~\ref{fig:cmd} 
HD\,133880\,B is a $1.15\,M_\odot$ pre-MS star at $\sim 16$\,Myrs, and therefore a 
likely member of the UCL for which \citey{deZeeuw99.1} give an age of $13$\,Myrs.  
As discussed above the B-type primary is a confirmed UCL member and probably younger 
than suspected by \citey{Hubrig01.1} who derived an 
age of $150$\,Myrs for HD\,133880\,A. We conclude that HD\,133880\,B is most probably a
true physical companion. 
\end{itemize}

\begin{table}
\begin{center}
\caption{Stellar parameters for {\em Chandra} observed IR sources near B-type stars. Mass, age, bolometric luminosity, and effective temperature are derived by comparison of published IR magnitudes and colors with evolutionary tracks from \protect\citey{Baraffe98.1}. The references (Col.~2) refer to the $JK$ magnitudes.}
\label{tab:cmd}
\begin{tabular}{lrrrrr}
\noalign{\smallskip}\hline\noalign{\smallskip}
HD        & Ref.     &  \multicolumn{1}{c}{$M$} & \multicolumn{1}{c}{$t$} & $\lg{L_{\rm bol}}$ & $T_{\rm eff}$ \\
          & ${\rm JK\,mag}$ & [$M_\odot$] & [Myrs] & [erg/s]           & [K] \\
\noalign{\smallskip}\hline\noalign{\smallskip}
113703    & (1)  & $0.9$  & $50$ & $33.27$ & $5020$ \\   
123445\,C & (2)  & $0.6$  & $4$  & $33.10$ & $3603$ \\
123445\,D & (2)  & $1.0$  & $10$ & $33.23$ & $4011$ \\
133880    & (3)  & $1.15$ & $16$ & $33.42$ & $4515$ \\
\noalign{\smallskip}\hline\noalign{\smallskip}
\multicolumn{6}{l}{(1) - \citey{Shatsky02.1}, (2) - \citey{Huelamo01.1},} \\
\multicolumn{6}{l}{(3) - \citey{Hubrig01.1} } \\
\end{tabular}
\end{center}
\end{table}

\section{Secondaries of Lindroos systems}\label{sect:lindroos}

As mentioned in the introduction two of the surveyed stars had previously known
late-type companions discovered by \citey{Lindroos86.1} in a photometric study of visual
binary stars, hereafter termed 'Lindroos systems'.
HD\,123445\,AB was observed and resolved with the {\em ROSAT} HRI, 
but the late-type companion was not detected (\cite{Huelamo00.1}). 
As also shown by \citey{Huelamo00.1} for HD\,113703 the {\em ROSAT} HRI provided
one elongated X-ray source, and left open whether this source corresponds to 
the primary or the secondary.
The present {\em Chandra} study separates the Lindroos binaries of 
HD\,113703 and HD\,123445.

Only one of the late-type components in these systems is
detected, HD\,113703\,B. We list its X-ray luminosity and the 95\,\% confidence 
upper limit for HD\,123445\,B in Table~\ref{tab:lindroos} 
together with optical parameters of interest. The upper limit was calculated
as described in Sect.~\ref{sect:lx}. 
The bolometric luminosities were computed with the 
bolometric corrections of \citey{Schmidt-Kaler82.1}. We assumed that both Lindroos 
secondaries are bound to the primaries, i.e. they are at the same distance as the 
B-type stars. 

Both Lindroos systems of Table~\ref{tab:lindroos} had been included in the optical 
spectroscopic sample examined by \citey{Pallavicini92.1}. The aim of that study was
to establish or refute the pre-MS nature of the presumed companions. 
Interestingly, only HD\,113703\,B was found to exhibit strong Lithium absorption 
(indicative of youth) as well as strong Ca\,II emission and filled-in H$\alpha$ profile
(both indicative of magnetic activity). HD\,123445\,B showed no sign for
a Lithium feature and neither Ca\,II nor H$\alpha$ activity, in agreement with its
non-detection in X-rays.

Ages for the late-type stars in Lindroos systems have been computed by \citey{Gerbaldi01.1}, 
who have studied the position of these objects
in the HR diagram using various sets of pre-MS models. 
We use the IR photometry for the Lindroos secondaries in the {\em Chandra}
sample in conjunction with the \citey{Baraffe98.1} IR CMD  
to check their evolutionary stage. 
For HD\,123445\,B \citey{Lindroos83.1} derive
$J-K=1.02 \pm 0.25$\,mag. 
Its photometric errors do not exclude that it is a pre-MS star. But its age 
should be younger than $\sim 10$\,Myrs. \citey{Gerbaldi01.1} have derived a similar age 
($3...8$\,Myrs). 
While \citey{Gerbaldi01.1} have argued that this Lindroos system is probably
not bound, we think that the distance and age of the B-type primary are not well
enough constrained to rule out a physical connection. 
No near-IR photometry is available for HD\,113703\,B. 
However, this star is most probably bound to its primary, given that
(i) they form a common radial velocity pair, (ii) it shows the Li I
absorption line in its spectrum, (iii)
its age is compatible with that of the primary (\cite{Gerbaldi01.1}), 
and (iv) it is a strong X-ray source. 

\begin{table}
\begin{center}
\caption{Secondaries of Lindroos systems observed with {\em Chandra}. Column 'Ref' refers to the separation and position angle.}
\label{tab:lindroos}
\begin{tabular}{llrrrlrr}
\noalign{\smallskip}\hline\noalign{\smallskip}
HD        & SpT    & \multicolumn{1}{c}{$V$}   &  Sep.  & P.A.  & Ref. & $\lg{L_{\rm bol}}$     & $\lg{L_{\rm x}}^*$ \\
          &        & \multicolumn{1}{c}{[mag]} & [$^{\prime\prime}$] & [$^\circ$] & & [erg/s] & [erg/s] \\
\noalign{\smallskip}\hline\noalign{\smallskip}
113703    & K0\,Ve & $11.5$ & $11.5$ & $79$  & (1) & $33.20$ & $30.3$  \\
123445    & K2\,V  & $13.0$ & $28.6$ & $35$  & (2) & $33.17$ & $<28.5$ \\
\noalign{\smallskip}\hline\noalign{\smallskip}
\multicolumn{8}{l}{(1) - Hu\'elamo et al., in prep., (2) - \citey{Turon93.1},} \\ 
\multicolumn{8}{l}{$^*$ in the $0.5-8$\,keV band} \\
\end{tabular}
\end{center}
\end{table}

\section{Summary}\label{sect:discussion}

In a sample of five late B-type stars with close but spatially resolved 
companion candidates discovered in AO observations {\em Chandra} 
observations have revealed all but one of the new IR objects 
as X-ray emitters,  
with $L_{\rm x}$ between $10^{29}$ and $10^{30}$\,erg/s.  
The only one which is not detected is the companion of HD\,169978. However,
the ACIS image displays an enhanced count rate over the local
background, and thus HD\,169978\,B may be a weaker X-ray source. 

Two of the late B-type primaries, HD\,1685 and HD\,169978, are detected. 
This could be due to either (i) further (even
closer) late-type companions not resolved in the AO observations or
(ii) to intrinsic X-ray emission from the late B-type stars.  
As mentioned in Sect.~\ref{sect:sample} \citey{Aerts99.1} found that HD\,169978 
is a single-lined spectroscopic binary.  
The X-ray luminosities of both HD\,1685\,A and HD\,169978\,A are comparable 
to those of the new IR objects, favoring low-mass companions as the site
of the X-ray production.  
The upper limits for the undetected B-type stars range between 
$\lg{L_{\rm x, lim}} \sim 27.8...28.5$\,erg/s
(dependent on the background emission level, exposure time, and distance of
the target). Most of the observations were short, but HD\,113703\,A is not 
detected despite the exposure was deeper.

We examined whether the new IR objects are true companions to the B-type
stars or just
chance projections by placing them on the CMD and comparing them to evolutionary
tracks. In three cases (HD\,113703\,C, HD\,123445\,CD, and HD\,133880\,B) 
the ages of the IR sources derived from the tracks are
compatible with them being on the pre-MS, and since the primary B-type stars are
known to be young these objects likely form
bound systems. However, definite confirmation of their status requires spectral
information that will prove or reject the youth of these companion candidates. 

The X-ray luminosities we measured with {\em Chandra} for our targets  
tend to be smaller than the 
values given before based on {\em ROSAT} data. We believe that the {\em ROSAT} values
are less reliable because they are just estimates based on hardness ratios,
while our observations with ACIS represent the first X-ray spectra for these stars.  
In contrast to an earlier conjecture by 
\citey{Berghoefer96.1} the spectra of this sample are not soft, but rather 
hard with most of the emission emanating at energies $> 0.5$\,keV. 
This is another indication for the youth of the objects because the strength
and hardness of the X-ray emission is known to decrease rapidly with
stellar age (e.g. \cite{Damiani95.1}, \cite{Stelzer01.1}). 

In terms of $L_{\rm x}/L_{\rm bol}$ the late-type stars, i.e. the Lindroos secondaries
and the new IR objects, are typical for young late-type stars: 
$\lg{(L_{\rm x}/L_{\rm bol})} \sim -3...-4$ for all of the detections.  
Strong X-ray emission points at youth, and given the youth of the primaries 
may be interpreted as indication that the objects form  
truely physically bound systems. The upper limit for
the Lindroos secondary to HD\,123445\,A places it at the lower end of the
typical activity range, consistent with the lack of youth signatures in its optical 
spectrum. Note that the upper limit to $L_{\rm x}$
for this object is uncertain because it is possibly 
at a different distance than the one assumed. 
For the companion to HD\,169978 a deeper observation is needed to obtain a useful 
constraint on its X-ray emission. 

Our observations support the trend of the primaries 
to split in two groups in the $L_{\rm x}/L_{\rm bol}-$ diagram. 
Such a bifurcation was recently pointed out by \citey{Daniel02.1} for 
A...F-type Pleiades stars: 
Apparent X-ray emitters on the one side, 
and on the other side X-ray quiet stars with upper limits by $1-2$ orders 
of magnitudes lower than the $L_{\rm x}/L_{\rm bol}$ values of the detections.  
In view of the fact that one of our two detected B-type stars and a few of the
Pleiades A...F-type stars are known to have a close companion unresolved even with
{\em Chandra} it seems reasonable to think  
that in the active group the X-rays are actually produced by the late-type companions,
which would move the objects up and to the left in Fig.~\ref{fig:acis_lx_lbol}. 
However, firm conclusions can only be drawn if it can be established that all of
these stars have faint companions. 

In our future work we will continue to track down the problem of X-ray emission 
from B-type stars by
(a) pushing upper limits on $L_{\rm x}/L_{\rm bol}$ well below the present values with help
of deeper X-ray observations, and
(b) examining the multiplicity of the apparently X-ray active B-type stars through 
IR spectroscopy. Both techniques combined on a large sample are likely to show that 
with present-day X-ray and IR instrumentation it is possible to approach a 
solution to this longstanding mystery.

\begin{acknowledgements}
We would like to thank D. Le Mignant for kindly providing us the AO images
of the objects discussed here, and an anonymous referee for helpful comments 
that improved the manuscript.  
BS acknowledges financial support from the European Union by the Marie
Curie Fellowship Contract No. HPMD-CT-2000-00013. 
This research has made use of the SIMBAD database, operated at CDS, Strasbourg, France,
and the {\em Hipparcos} catalogue accessed through the VizieR data base. 
\end{acknowledgements}

\end{document}